\documentclass[aps,twocolumn]{revtex4}

\RequirePackage[T2A]{fontenc}

\usepackage{amssymb}
\usepackage{amsmath}
\usepackage{bm}
\usepackage[dvips]{epsfig}
\usepackage{graphicx}
\usepackage{textcomp}

\begin{document}

\title{On chaotic regimes of conductivity behavior in the 
tight-binding approximation} 

\author{A.Ya. Maltsev}

\affiliation{
\centerline{\it L.D. Landau Institute for Theoretical Physics}
\centerline{\it 142432 Chernogolovka, pr. Ak. Semenova 1A,
maltsev@itp.ac.ru}}

\begin{abstract}
 We investigate the probability of detecting the most nontrivial 
conductivity behavior regimes in metals whose electron spectrum 
is described by the tight-binding approximation. These regimes 
are associated with the emergence of highly complex electron 
trajectories on the Fermi surface and correspond to a nontrivial 
(scaling) behavior of the conductivity tensor in strong magnetic 
fields. The geometry of such trajectories, as well as the 
corresponding conductivity regimes, have been well studied 
theoretically; however, they have not yet been observed experimentally. 
The results of our study allow us, in particular, to estimate the 
probability of their occurrence and to indicate the conditions for 
their possible detection for a wide class of conductors.
\end{abstract}

\maketitle

\section{Introduction}

 In this paper, we investigate the probability of the occurrence 
of special ``chaotic'' regimes of conductivity behavior in strong 
magnetic fields in conductors with an electron spectrum given
by the tight-binding approximation. More precisely, we consider 
the conductivity in high-purity (single-crystal) samples at low 
temperatures ($T \, \leq \, 1 {\rm K}$) in magnetic fields 
$B \, \geq \, 1 {\rm Tl} \, $. This situation corresponds to a large 
electron mean free path, ensuring the fulfillment of the condition 
$\, \omega_{B} \tau \gg 1 \, $, where 
$\, \omega_{B} \, = \, e B / m^{*} c \, $ is the cyclotron frequency, 
and $\, \tau \, $ is the electron mean free time in the crystal.

 As is well known 
(see e.g. \cite{Kittel, Ziman, AshcroftMermin, Abrikosov}), 
the semiclassical dynamics of an electron in external electric 
and magnetic fields is given by the system
\begin{equation}
\label{rsyst}
{\dot {\bf r}} \,\,\, = \,\,\, {\bf v}_{\rm gr} ({\bf p})
\,\,\, \equiv \,\,\, \nabla \epsilon ({\bf p})
\end{equation}
\begin{equation}
\label{psyst}
{\dot {\bf p}} \,\,\, = \,\,\, e {\bf E} \,\, + \,\, 
{e \over c} \,\,
\left[ {\bf v}_{\rm gr} ({\bf p}) \times {\bf B} \right]
\,\,\, \equiv \,\,\, e {\bf E} \,\, + \,\, 
{e \over c} \,\,
\left[ \nabla \epsilon ({\bf p}) \times {\bf B} \right]
\end{equation}

 In system (\ref{rsyst}) - (\ref{psyst}), the vector
$\, {\bf r} \, $ represents the electron coordinate in a crystal, 
while $\, {\bf p} \, $ is the quasi-momentum of the electron.

 The quantity $\, {\bf p} \, $ is defined modulo the vectors of 
the reciprocal lattice $\, L^{*} \, $, whose basis is given by the 
vectors
\begin{multline}
\label{a1a2a3}
{\bf a}_{1} \,\,\, = \,\,\, 2 \pi \hbar \,\,
{{\bf l}_{2} \, \times \, {\bf l}_{3} \over
({\bf l}_{1}, \, {\bf l}_{2}, \, {\bf l}_{3} )} \,\,\, , \quad \quad
{\bf a}_{2} \,\,\, = \,\,\, 2 \pi \hbar \,\,
{{\bf l}_{3} \, \times \, {\bf l}_{1} \over
({\bf l}_{1}, \, {\bf l}_{2}, \, {\bf l}_{3} )} \,\,\, ,  \\
{\bf a}_{3} \,\,\, = \,\,\, 2 \pi \hbar \,\,
{{\bf l}_{1} \, \times \, {\bf l}_{2} \over
({\bf l}_{1}, \, {\bf l}_{2}, \, {\bf l}_{3} )} 
\quad \quad \quad \quad
\end{multline}
(where $\, ({\bf l}_{1}, {\bf l}_{2}, {\bf l}_{3}) \, $ 
define the basis of the crystallographic lattice $\, L$).

 In other words, we can say that the quasi-momentum 
$\, {\bf p} \, $ is a point of the three-dimensional torus 
$\, \mathbb{T}^{3} \, $, defined by the factorization of 
the space $\, \mathbb{R}^{3} \, $ with respect to the 
reciprocal lattice vectors
$$\mathbb{T}^{3} \,\,\, = \,\,\, \mathbb{R}^{3} \Big/  L^{*} $$

 The dependence of the electron energy $\, \epsilon ({\bf p}) \, $ 
on its quasi-momentum (dispersion relation) can also be considered 
either as a function on the torus $\, \mathbb{T}^{3} \, $ or as 
a 3-periodic function in $\, \mathbb{R}^{3} \, $ with 
periods (\ref{a1a2a3}).

 As can be easily seen from (\ref{rsyst}) - (\ref{psyst}), 
the electron trajectories in $\, {\bf r}$ - space are completely 
determined by the solutions of the subsystem (\ref{psyst}) 
(and the initial conditions), i.e. by the electron trajectories 
in $\, {\bf p}$ - space.

 In our situation, the value of $\, {\bf E} \, $ in the 
system (\ref{psyst}) is assumed to be infinitely small. At the 
same time, the condition $\, \omega_{B} \tau \gg 1 \, $ assumes 
a rather large value of the magnetic field. As a consequence, 
the main role in the description of transport phenomena in the 
limit $\, \omega_{B} \tau \gg 1 \, $ is actually played 
by the system
\begin{equation}
\label{MFSyst}
{\dot {\bf p}} \,\,\,\, = \,\,\,\, {e \over c} \,\,
\left[ {\bf v}_{\rm gr} ({\bf p}) \times {\bf B} \right]
\,\,\,\, \equiv \,\,\,\, {e \over c} \,\, \left[ \nabla \epsilon ({\bf p})
\times {\bf B} \right] 
\end{equation}

\begin{figure}[t]
\begin{center}
\includegraphics[width=\linewidth]{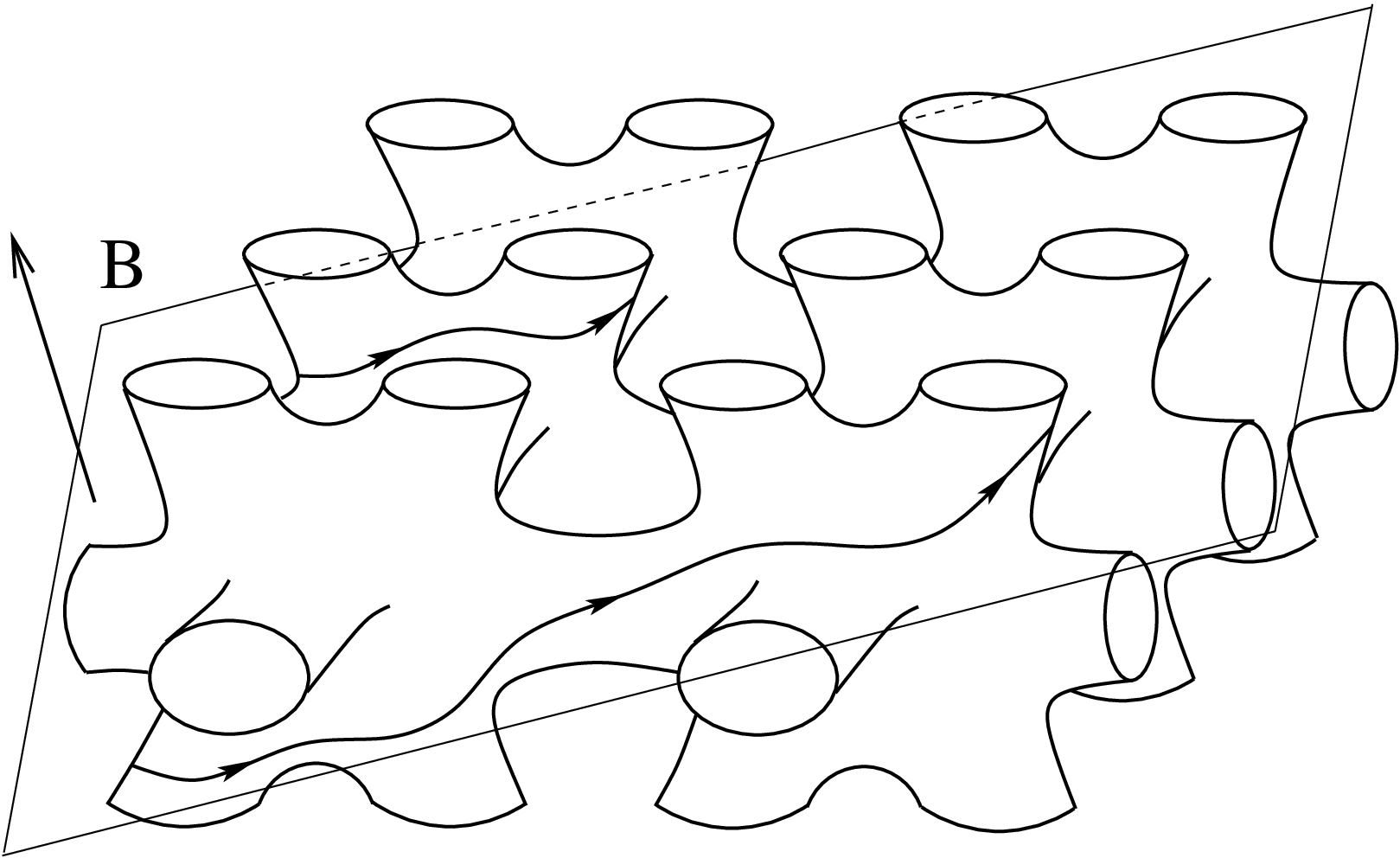}
\end{center}
\caption{Trajectories of system (\ref{MFSyst}) 
in the $\, {\bf p}$ - space.}
\label{ArbSurf}
\end{figure}

 The trajectories of system (\ref{MFSyst}) in the full 
$\, {\bf p}$ - space are defined by the intersections of planes 
orthogonal to $\, {\bf B} \, $, and the 3 - periodic surfaces 
$\, \epsilon ({\bf p}) = {\rm const} \, $ (Fig. \ref{ArbSurf}).

 As we have already said, the electron trajectories 
$\, {\bf r} ({\bf t}) \, $ in coordinate space are completely 
determined by the corresponding solutions 
$\, {\bf p} ({\bf t}) \, $ of the system (\ref{psyst}). 
When passing to the system (\ref{MFSyst}), one can also see 
that the projections of the trajectories 
$\, {\bf r} ({\bf t}) \, $ onto the plane orthogonal 
to $\, {\bf B} \, $ become similar to the trajectories 
$\, {\bf p} ({\bf t}) \, $, rotated by $\, 90^{\circ}$.

 As can be seen (Fig. \ref{ArbSurf}), the trajectories of 
the system (\ref{MFSyst}) can be either closed or unclosed 
(open) in $\, {\bf p}$ - space. The shape of the trajectories 
(\ref{MFSyst}) depends on both the shape of the surface 
$\, \epsilon ({\bf p}) = {\rm const} \, $ and the direction 
of the magnetic field.

 As was first shown in the I.M. Lifshitz school in the 1950s 
(see \cite{lifazkag,lifpes1,lifpes2,etm}), the behavior of 
conductivity (and other transport phenomena) in a metal at 
$\, \omega_{B} \tau \gg 1 \, $ is determined mainly by the shape 
of the trajectories of system (\ref{MFSyst}) and, in particular, 
by the presence or absence of its open trajectories on the Fermi 
surface.

\begin{figure}[t]
\begin{center}
\includegraphics[width=\linewidth]{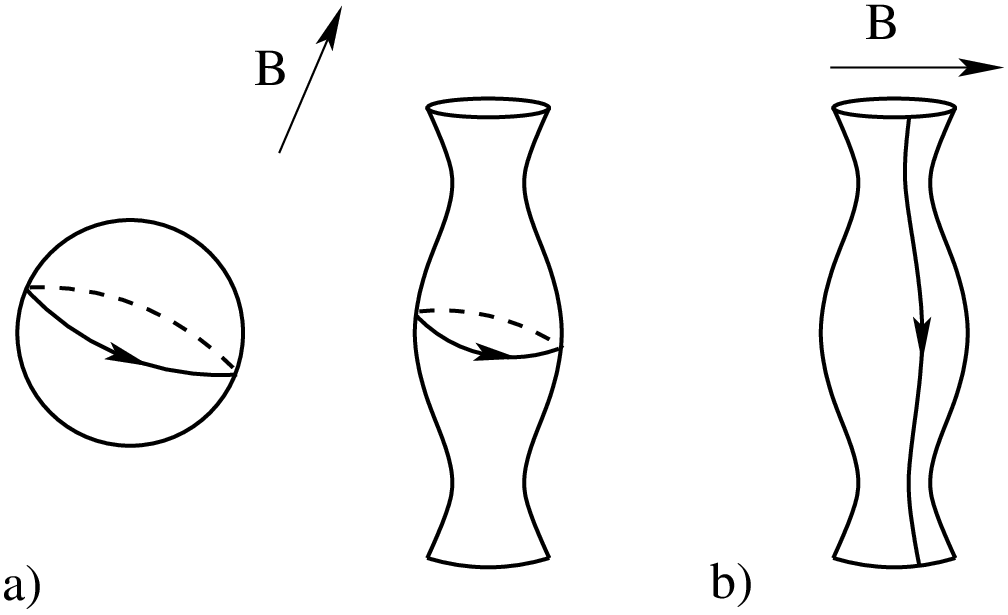}
\end{center}
\caption{Closed (a) and open periodic trajectories (b) of 
the system (\ref{MFSyst}) on the Fermi surface.}
\label{ClandPer}
\end{figure}

 In particular, the contribution of closed and open periodic 
trajectories (Fig. \ref{ClandPer}) to the conductivity tensor 
in the limit $\, \omega_{B} \tau \rightarrow \, \infty \, $ can be 
represented by the following asymptotic formulas (\cite{lifazkag})
\begin{equation}
\label{Closed}
\Delta \sigma^{kl}_{\rm closed} \,\,\,\, \simeq \,\,\,\,
{n e^{2} \tau \over m^{*}} \, \left(
\begin{array}{ccc}
( \omega_{B} \tau )^{-2}  &  ( \omega_{B} \tau )^{-1}  &
( \omega_{B} \tau )^{-1}  \cr
( \omega_{B} \tau )^{-1}  &  ( \omega_{B} \tau )^{-2}  &
( \omega_{B} \tau )^{-1}  \cr
( \omega_{B} \tau )^{-1}  &  ( \omega_{B} \tau )^{-1}  &  *
\end{array}  \right)  
\end{equation}
(closed trajectories),
\begin{equation}
\label{Periodic}
\Delta \sigma^{kl}_{\rm periodic} \,\,\,\, \simeq \,\,\,\,
{n e^{2} \tau \over m^{*}} \, \left(
\begin{array}{ccc}
( \omega_{B} \tau )^{-2}  &  ( \omega_{B} \tau )^{-1}  &
( \omega_{B} \tau )^{-1}  \cr
( \omega_{B} \tau )^{-1}  &  *  &  *  \cr
( \omega_{B} \tau )^{-1}  &  *  &  *
\end{array}  \right)  
\end{equation}
(open periodic trajectories).

 The formulas (\ref{Closed}) and (\ref{Periodic}) describe 
the asymptotic behavior of the tensor $\, \sigma^{kl} (B) \, $ 
in the limit $\, \omega_{B} \tau \rightarrow \, \infty \, $. 
In particular, it is assumed that each matrix element 
in (\ref{Closed}) - (\ref{Periodic}) is defined, in fact, 
up to a constant factor, and the symbol $\, * \, $ denotes 
some dimensionless constant of order 1.

 In formulas (\ref{Closed}) - (\ref{Periodic}), the direction 
of the $\, z \, $ axis coincides with the direction of the 
magnetic field. Moreover, in formula (\ref{Periodic}), the 
direction of the $\, x \, $ axis coincides with the mean 
direction of periodic trajectories in the $\, {\bf p}$ - space. 
As can be seen, formula (\ref{Periodic}) decribes a 
strong anisotropy of conductivity in the plane orthogonal 
to $\, {\bf B} \, $ in the limit 
$\, \omega_{B} \tau \rightarrow \, \infty \, $. 

 The quantities $\, n \, $ and $\, m^{*} \, $ represent 
the concentration and effective mass of electrons in the 
crystal and are also defined by order of magnitude. It can be 
noted that the quantity 
$\, \omega_{B} \, = \, e B / m^{*} c \, $ here also 
has a conventional meaning, and the condition 
$\, \omega_{B} \tau \gg 1 \, $ actually means that 
the electron travels a sufficiently large distance 
(noticeably greater than the size of the Brillouin zone) 
along the trajectories of system (\ref{MFSyst}) between 
two scattering acts. Taking into account the above remarks, 
formulas (\ref{Closed}) - (\ref{Periodic}) well describe 
the behavior of the tensor $\, \sigma^{kl} (B) \, $ in 
both the presented situations.

 Periodic trajectories represent the simplest open 
trajectories of system (\ref{MFSyst}). However 
(see, e.g., \cite{lifpes1,lifpes2,etm}), they are not 
the only type of open trajectories of (\ref{MFSyst}). 
It should be noted, however, that more complex open 
trajectories of the system (\ref{MFSyst}) can arise 
only on Fermi surfaces of sufficient complexity 
(Fig. \ref{ArbSurf}).

 The problem of complete classification of open trajectories 
of the system (\ref{MFSyst}) for arbitrary dispersion laws 
$\, \epsilon ({\bf p}) \, $ was set by S.P. Novikov in 
\cite{MultValAnMorseTheory}. The study of this problem in 
his topological school 
(\cite{zorich1,dynn1992,Tsarev,dynn1,zorich2,DynnBuDA,dynn2,dynn3}) 
led to a number of highly nontrivial topological results, providing 
the basis for the complete classification of various types of open 
trajectories of (\ref{MFSyst}), obtained to date.

 In a sense, the ``main'' type of open trajectories of system 
(\ref{MFSyst}) is represented by stable open trajectories, 
i.e., trajectories that are stable with respect to small variations 
in the dispersion relation $\, \epsilon ({\bf p}) \, $, as well as 
the direction of $\, {\bf B} \, $. In this paper, such trajectories 
will not be the subject of an independent study. However, we will 
need related structures to consider more complex trajectories of 
(\ref{MFSyst}). Therefore, we give here a brief description of 
the general properties of such trajectories.

\begin{figure}[t]
\begin{center}
\includegraphics[width=\linewidth]{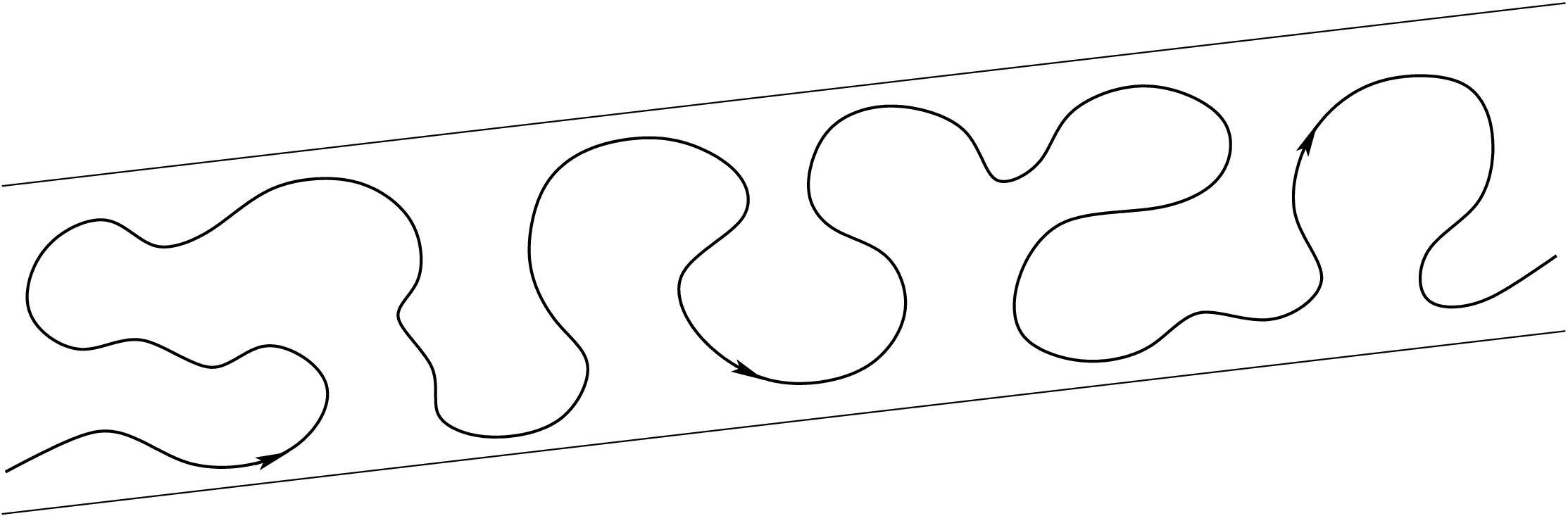}
\end{center}
\caption{The form of a stable open trajectory of system 
(\ref{MFSyst}) in a plane orthogonal to 
$\, {\bf B} \, $ (schematically).}
\label{StableTr}
\end{figure}

  First of all, the stable open trajectories of system 
(\ref{MFSyst}) possess remarkable geometric properties. 
Namely (\cite{zorich1,dynn1992,dynn1}), each stable 
open trajectory of (\ref{MFSyst}) lies in a straight strip of 
finite width (in each of the planes orthogonal to $\, {\bf B}$), 
passing through it (Fig. \ref{StableTr}). Note that stable open 
trajectories of (\ref{MFSyst}) are generally not periodic.

 The second important property of stable open trajectories 
of system (\ref{MFSyst}) is that their mean direction 
in $\, {\bf p}$ - space (in all planes orthogonal to $\, {\bf B}$) 
is always given by the intersection of the plane orthogonal 
to $\, {\bf B} $ and some fixed integer (generated by two 
reciprocal lattice vectors) plane $\, \Gamma \, $, which 
remains unchanged under small variations in the direction 
of $\, {\bf B} $ and the energy level $\, \epsilon \, $. 
This property can be observed experimentally and was the 
basis for the introduction of ``topological numbers'' observed 
in the conductivity of normal metals in \cite{PismaZhETF}.  

 In general, each dispersion relation $\, \epsilon ({\bf p}) \, $ 
is characterized by its own set of domains $\, V_{\alpha} \, $ 
in the space of parameters $\, {\bf B} \, $ and $\, \epsilon \, $, 
corresponding to the emergence of stable open trajectories 
associated with some integer planes $\, \Gamma_{\alpha} \, $.

 The set of domains $\, V_{\alpha} \, $ for a fixed dispersion 
relation $\, \epsilon ({\bf p}) \, $ admits a rather convenient 
description since the emergence of stable open trajectories at 
some level $\, \epsilon ({\bf p}) \, = \, \epsilon_{0} \, $ 
for a fixed direction $\, {\bf B} \, $ admits the emergence 
of only stable open trajectories of the same direction also 
at all other levels $\, \epsilon ({\bf p}) \, = \, \epsilon \, $.

 Therefore, instead of the parameters 
$\, \left( {\bf B}, \epsilon \right) \, $ we can restrict ourselves 
to the parameter
$${\bf n} \,\,\, = \,\,\, {\bf B} / B \,\,\, \in \,\,\,
\mathbb{S}^{2} \,\,\, , $$
i.e., only to the direction of the magnetic field.

 In fact, a stronger statement holds (\cite{dynn3}). Namely, 
the emergence of open trajectories for a given direction 
of $\, {\bf B} \, $ always occurs in some closed energy interval
$$\epsilon \,\,\, \in \,\,\, \left[ \epsilon_{1} ({\bf B}) , \,
\epsilon_{2} ({\bf B}) \right] \,\,\, , $$
which can be contracted to a point
$$\epsilon_{0} ({\bf B}) \,\,\, = \,\,\, \epsilon_{1} ({\bf B})
\,\,\, = \,\,\, \epsilon_{2} ({\bf B}) $$
(all trajectories for
$\, \epsilon \,\,\, \notin \,\,\, \left[ \epsilon_{1} ({\bf B}) , \,
\epsilon_{2} ({\bf B}) \right] \, $ are closed).

\begin{figure}[t]
\begin{center}
\includegraphics[width=\linewidth]{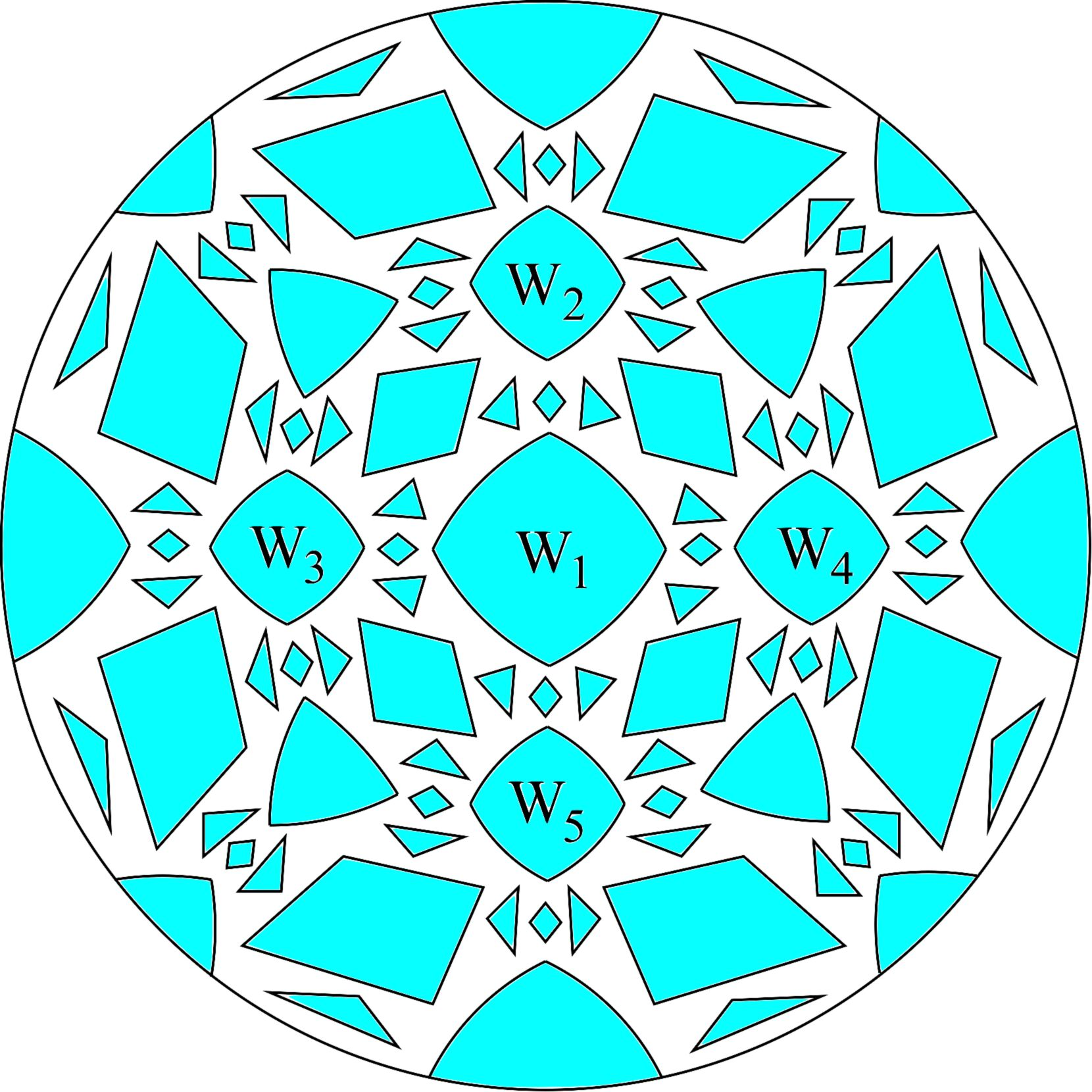}
\end{center}
\caption{The angular diagram of a fixed dispersion relation 
$\, \epsilon ({\bf p}) \, $ (schematically).}
\label{FullDiagram}
\end{figure}

 Instead of the domains $\, V_{\alpha} \, $, it is then sufficient 
to specify domains $\, W_{\alpha} \subset \mathbb{S}^{2} \, $ 
corresponding to the emergence of stable open trajectories of 
(\ref{MFSyst}) (at least at one level $\, \epsilon $) and define 
the functions $\, \epsilon_{1} ({\bf n}) \, $ and 
$\, \epsilon_{2} ({\bf n}) \, $, as well as the plane 
$\, \Gamma_{\alpha} \, $, in each of these domains 
(Fig. \ref{FullDiagram}).

 We will call here the domains 
$\, W_{\alpha} \subset \mathbb{S}^{2} \, $ ``Stability Zones'', 
and their location on the sphere - the angular diagram of the 
dispersion relation $\, \epsilon ({\bf p}) \, $.

 As shown in \cite{dynn3}, the Stability Zones $\, W_{\alpha} \, $ 
represent domains with piecewise smooth boundaries, and their union 
is everywhere dense on the sphere $\, \mathbb{S}^{2} \, $. 
Moreover, for relations $\, \epsilon ({\bf p}) \, $ satisfying 
the ``physical'' condition
$$\epsilon ({\bf p}) \,\,\, = \,\,\, \epsilon (- {\bf p}) \,\,\, , $$
the union of all $\, W_{\alpha} \, $ forms a set of the full 
measure on $\, \mathbb{S}^{2} \, $ (I.A. Dynnikov, P. Hubert, 
P. Mercat, A.S. Skripchenko, in preparation).

 The functions $\, \epsilon_{1} ({\bf n}) \, $ and 
$\, \epsilon_{2} ({\bf n}) \, $ are continuous on the set of 
directions of $\, {\bf B} \, $ that do not lead to periodic 
open trajectories of system (\ref{MFSyst}). In particular, 
they are continuous on the set of directions of $\, {\bf B} \, $ 
of maximal irrationality, so they can be extended to continuous 
functions $\, \widetilde{\epsilon}_{1} ({\bf n}) $, 
$\, \widetilde{\epsilon}_{2} ({\bf n}) \, $, defined everywhere 
on the unit sphere $\, \mathbb{S}^{2} \, $. For special directions 
of $\, {\bf B} \, $, leading to the emergence of periodic open 
trajectories of (\ref{MFSyst}), the functions 
$\, \epsilon_{1} ({\bf n}) $, $\, \epsilon_{2} ({\bf n}) \, $ 
have jumps, such that
$$\epsilon_{1} ({\bf n}) \,\,\, \leq \,\,\, 
\widetilde{\epsilon}_{1} ({\bf n}) \,\,\, \leq \,\,\,
\widetilde{\epsilon}_{2} ({\bf n}) \,\,\, \leq \,\,\,
\epsilon_{2} ({\bf n}) $$
(see \cite{dynn3}).

 It is interesting that all angular diagrams (for arbitrary 
$\, \epsilon ({\bf p}) $) can be divided into two types, namely, 
diagrams containing only one Stability Zone (covering the entire 
sphere $\, \mathbb{S}^{2} $) and diagrams containing an infinite 
number of Zones $\, W_{\alpha} \, $ (\cite{dynn3}). Diagrams of 
the first type arise for some specific (mainly quasi-one-dimensional) 
classes of conductors, and we will not consider them here. For most 
conductors, the corresponding diagrams contain an infinite number 
of Stability Zones $\, W_{\alpha} \, $. In particular, this applies 
to the model dispersion relations $\, \epsilon ({\bf p}) \, $ 
considered here.

 As can be seen, stable open trajectories of system (\ref{MFSyst}) 
arise for almost all (or all) directions of $\, {\bf B} \, $ for 
any dispersion relation $\, \epsilon ({\bf p}) \, $. At the same time, 
periodic and stable open trajectories do not exhaust all the types 
of open trajectories of (\ref{MFSyst}) (\cite{Tsarev,DynnBuDA,dynn2}). 
Trajectories of other types have much more complex behavior and we 
will call them chaotic.

 The first example of a chaotic trajectory of system (\ref{MFSyst}) 
was constructed by S.P. Tsarev (\cite{Tsarev}). Tsarev-type trajectories 
have an asymptotic direction in planes orthogonal to $\, {\bf B} \, $, 
but cannot be contained in any straight strip of finite width. 
The contribution of Tsarev-type trajectories to the conductivity 
tensor differs only slightly from the contribution (\ref{Periodic}); 
in particular, it also corresponds to strong anisotropy of conductivity 
in the plane orthogonal to $\, {\bf B} \, $. Tsarev-type trajectories 
represent one of the main types of chaotic trajectories of (\ref{MFSyst}) 
(according to the classification) and describe chaotic trajectories that 
arise for ``partially'' irrational directions of $\, {\bf B} \, $ 
(the plane orthogonal to $\, {\bf B} \, $ contains one reciprocal 
lattice vector).

\begin{figure}[t]
\begin{center}
\includegraphics[width=\linewidth]{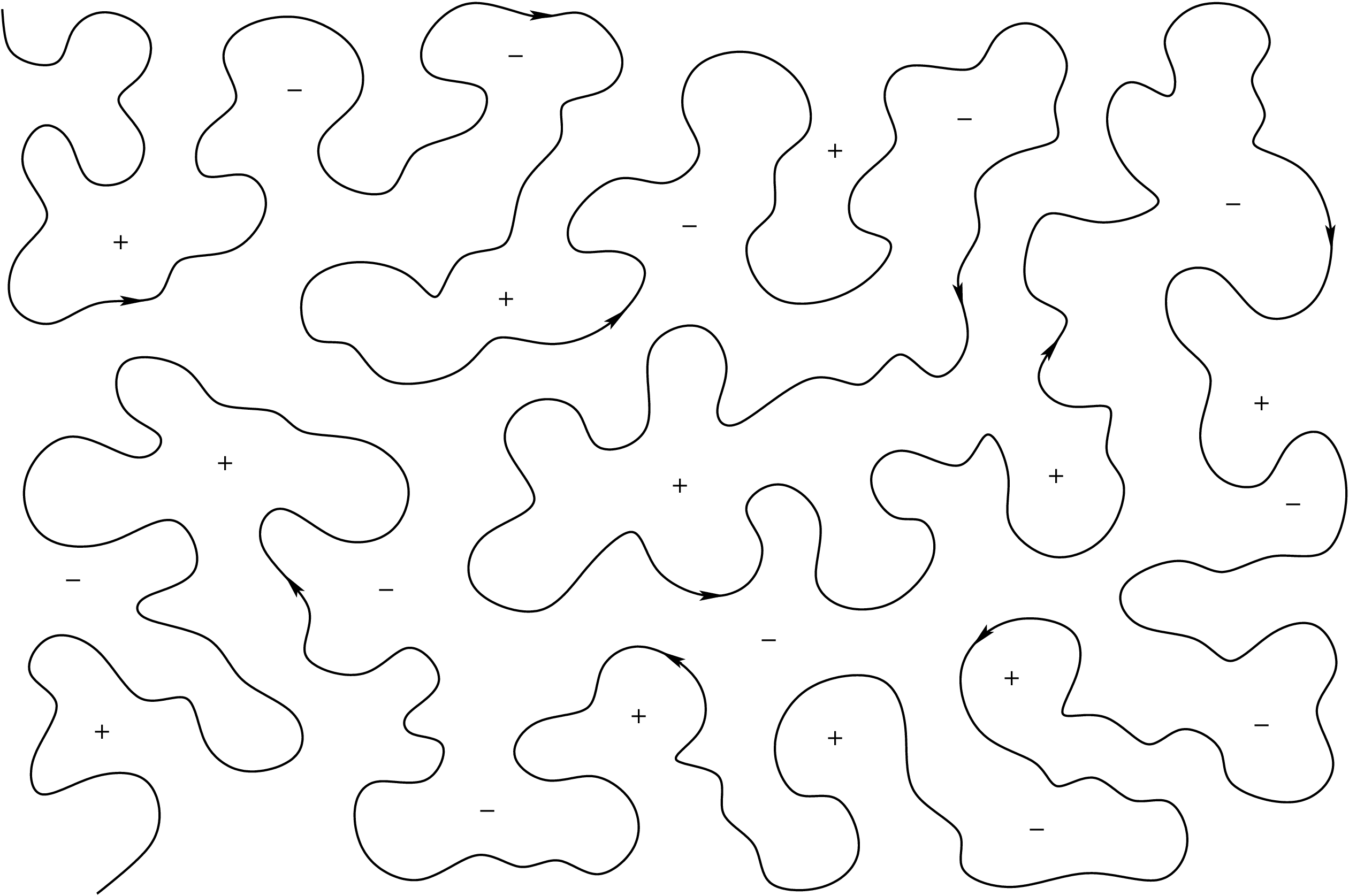}
\end{center}
\caption{The form of Dynnikov's chaotic trajectory (schematically).}
\label{DynnTr}
\end{figure}

 Chaotic trajectories of a more complex type (Dynnikov's type) 
were discovered in the works \cite{DynnBuDA,dynn2}. Trajectories 
of this type can arise only for directions of $\, {\bf B} \, $ of 
maximal irrationality and have a very complex geometry, 
wandering ``everywhere'' in the plane orthogonal to $\, {\bf B} \, $ 
(Fig. \ref{DynnTr}).

 The contribution of Dynnikov's chaotic trajectories to the 
conductivity tensor is the most nontrivial (\cite{ZhETF2,TrMIAN}). 
One of the features of this contribution is the vanishing of all 
components of $\, \Delta \sigma^{kl}_{\rm chaotic} (B) \, $ 
(including the conductivity along the direction of $\, {\bf B}$) 
in the limit $\, \omega_{B} \tau \rightarrow \infty \, $:
$$\Delta \sigma^{kl}_{\rm chaotic} (B) \,\,\,\, = \,\,\,\,
{n e^{2} \tau \over m^{*}} \, \left(
\begin{array}{ccc}
o (1)  &  o (1)  &  o (1)  \cr
o (1)  &  o (1)  &  o (1)  \cr
o (1)  &  o (1)  &  o (1)
\end{array}  \right)  $$

 Another feature of this contribution is the emergence of 
fractional powers of $\, \omega_{B} \tau \, $ in the dependence 
of the components $\, \Delta \sigma^{kl}_{\rm chaotic} (B) \, $ 
on $\, B \, $. This behavior of 
$\, \Delta \sigma^{kl}_{\rm chaotic} (B) \, $ reflects a 
scaling behavior of Dynnikov's trajectories in the planes orthogonal 
to $\, {\bf B} \, $, which, as a rule, has anisotropic properties. 
With a suitable choice of axes $\, x \, $ and $\, y \, $ we can write 
for the components $\, \Delta \sigma^{kl}_{\rm chaotic} (B) \, $:
$$\Delta \sigma^{xx}_{\rm chaotic} (B) \,\,\, \simeq \,\,\,
{n e^{2} \tau \over m^{*}} \,
\left( \omega_{B} \tau \right)^{2\nu_{1}-2} \quad ,  $$
\begin{equation}
\label{ChaoticContr}
\Delta \sigma^{yy}_{\rm chaotic} (B) \,\,\, \simeq \,\,\,
{n e^{2} \tau \over m^{*}} \,
\left( \omega_{B} \tau \right)^{2\nu_{2}-2} \quad ,   
\end{equation}
$$\Delta \sigma^{zz}_{\rm chaotic} (B) \,\,\, \simeq \,\,\,
{n e^{2} \tau \over m^{*}} \,
\left( \omega_{B} \tau \right)^{2\nu_{3}-2} \quad , $$
where $\, 0 \, < \, \nu_{1}, \, \nu_{2}, \, \nu_{3} \, < \, 1 \, $ 
represent the main scaling indices of a trajectory.

 It should be noted that, in the general case, the 
contribution (\ref{ChaoticContr}) must be summed with the 
contribution (\ref{Closed}) given by closed trajectories on the 
Fermi surface. As a consequence, the conductivity 
$\, \sigma^{zz} (B) \, $ is suppressed only partially, 
remaining finite at $\, B \rightarrow \infty \, $. In the plane 
orthogonal to $\, {\bf B} \, $, the scaling contribution 
(\ref{ChaoticContr}) to the conductivity remains dominant.

\begin{figure}[t]
\begin{center}
\includegraphics[width=\linewidth]{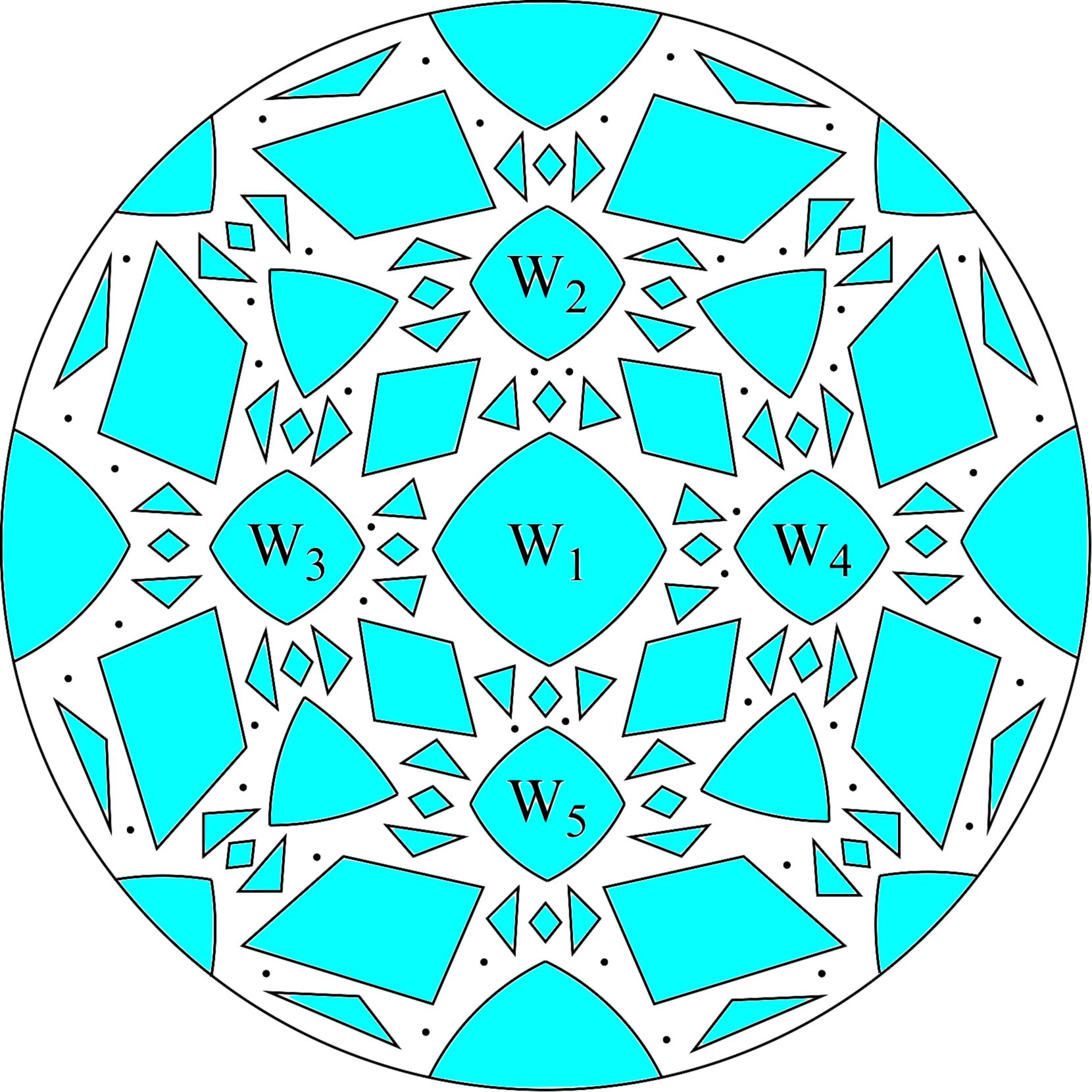}
\end{center}
\caption{``Chaotic'' directions of $\, {\bf B} \, $ on the 
angular diagram (schematically).}
\label{ChaoticDir}
\end{figure}

 All chaotic trajectories of system (\ref{MFSyst}) are 
unstable and, in particular, can arise only at a single level 
$\, \epsilon ({\bf p}) \, = \, \epsilon_{0} ({\bf n}) \, $ 
for a given direction of $\, {\bf B} \, $ (\cite{dynn1}). 
The corresponding directions of $\, {\bf B} \, $ lie in the 
complement $\, {\cal M} \, $ to the union of Zones 
$\, W_{\alpha} \, $ (together with the boundaries) 
on the unit sphere
$${\bf n} \quad \in \quad {\cal M} \,\,\, = \,\,\, 
\mathbb{S}^{2} \, \Big{\backslash} \, \bigcup_{\alpha} \,
\overline{W}_{\alpha} $$
(Fig. \ref{ChaoticDir}).

 The set $\, {\cal M} \, $ is a complex set of fractal type on 
$\, \mathbb{S}^{2} \, $. According to the conjecture of S.P. Novikov, 
its fractal dimension is strictly less than 2. In general, both the 
properties of the set $\, {\cal M} \, $ and the properties of 
chaotic trajectories are the subject of intensive research 
and are currently showing increasingly interesting features (see
\cite{zorich3,DeLeo1,DeLeo2,DeLeoPhysLettA,DeLeoPhysB,DeLeo3,
DeLeoDynnikov1,Dynnikov2008,DeLeoDynnikov2,Skripchenko1,Skripchenko2,
DynnSkrip1,DynnSkrip2,AvilaHubSkrip1,AvilaHubSkrip2,DynHubSkrip}).

\vspace{1mm}

 As can be seen, the search for chaotic trajectories of system 
(\ref{MFSyst}) is a rather complex problem and requires, in any 
case, a special choice of the direction of $\, {\bf B} \, $. 
At the same time, such trajectories necessarily arise for 
dispersion relations $\, \epsilon ({\bf p}) \, $, having
angular diagrams with an infinite number of Stability Zones 
$\, W_{\alpha} \, $ (\cite{dynn3}). Experimental 
observation of the contributions (\ref{ChaoticContr}), however, 
is significantly limited by the fact that they can only be made 
by trajectories lying on the Fermi surface 
$\, \epsilon ({\bf p}) \, = \, \epsilon_{F} \, $.

 The aim of this work is to estimate the probability of 
observing ``chaotic'' regimes (\ref{ChaoticContr}) in real conductors, 
based on numerical studies of the angular diagrams of real dispersion 
relations. As model dispersion relations, we use relations arising in 
the tight-binding approximation in crystals with cubic symmetry. 
The results of the numerical studies allow us not only to estimate 
the probability of detecting ``chaotic'' regimes but also to identify 
the conductor parameters for which their detection is most likely.

\section{Angular diagrams of Fermi surfaces and chaotic regimes}
\setcounter{equation}{0}

 As we have already said, the behavior of the tensor 
$\, \sigma^{kl} (B) \, $ is determined only by the trajectories of 
(\ref{MFSyst}) lying on the Fermi surface 
$$S_{F} \, : \quad \epsilon ({\bf p}) \,\,\, = \,\,\, \epsilon_{F} $$

 The emergence of stable open trajectories on the surface 
$\, S_{F} \, $ for a given direction of $\, {\bf B} \, $ 
is determined by two conditions
$${\bf n} \, \in \, W_{\alpha} \,\,\, , \quad \quad
\widetilde{\epsilon}_{1} ({\bf n}) \,\,\, \leq \,\,\, \epsilon_{F}
\,\,\, \leq \,\,\, \widetilde{\epsilon}_{2} ({\bf n}) $$

 Similarly, the emergence of chaotic trajectories on the Fermi 
surface requires an additional condition
$$\epsilon_{0} ({\bf n}) \,\,\, = \,\,\, \epsilon_{F} $$
for the corresponding direction of $\, {\bf B} \, $.

 For directions of $\, {\bf B} \, $, such that
$$ {\bf n} \, \in \, W_{\alpha} \,\,\, , \quad \quad 
\epsilon_{F} \,\,\, \notin \,\,\, 
\left[ \widetilde{\epsilon}_{1} ({\bf n}) , \,
\widetilde{\epsilon}_{2} ({\bf n}) \right] $$
or
$${\bf n} \, \in \, {\cal M} \,\,\, , \quad \quad 
\epsilon_{F} \,\,\, \neq \,\,\, \epsilon_{0} ({\bf n}) $$
the Fermi surface will contain only closed trajectories of 
(\ref{MFSyst}) or unstable periodic trajectories (for special 
directions of $\, {\bf B} \, $ of non-maximal irrationality).

 It is natural to consider now the angular diagrams indicating 
the type of open trajectories of (\ref{MFSyst}) (or their absence) 
on the Fermi surface. Such angular diagrams may, in particular, 
contain new Stability Zones $\, \Omega_{\alpha} \, $, 
corresponding to the emergence of stable open trajectories on the 
Fermi surface, as well as (the most interesting to us) directions 
of $\, {\bf B} \, $, corresponding to the emergence of chaotic 
trajectories on $\, S_{F} \, $. In addition, the new angular diagrams 
will contain domains (on $\, \mathbb{S}^{2}$) corresponding to the 
presence of only closed trajectories on $\, S_{F} \, $, as well as 
lines of directions of $\, {\bf B} \, $, corresponding to the 
emergence of unstable periodic trajectories on $\, S_{F} \, $.

\begin{figure}[t]
\begin{center}
\includegraphics[width=\linewidth]{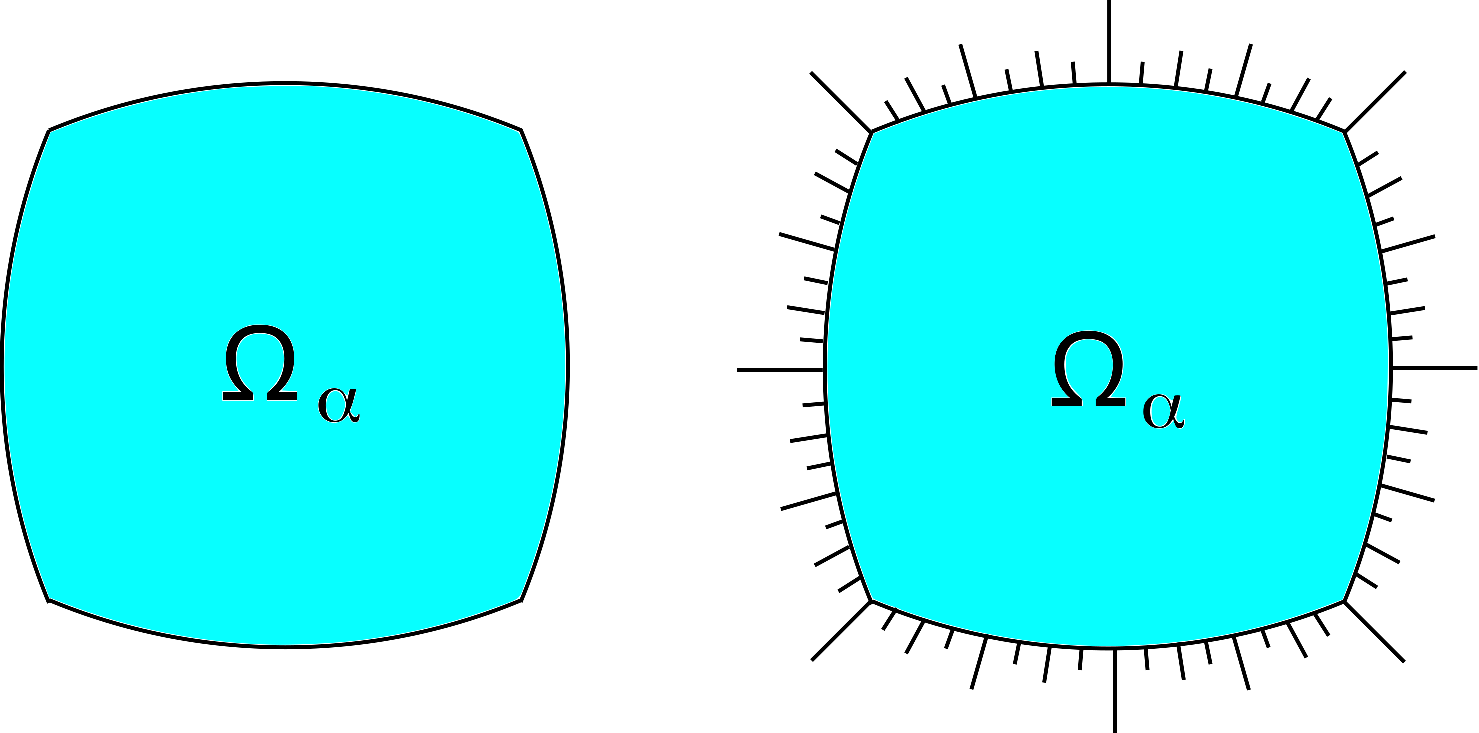}
\end{center}
\caption{A Stability Zone $\, \Omega_{\alpha} \, $ and the 
adjacent lines of directions of $\, {\bf B} \, $, corresponding 
to the emergence of unstable periodic trajectories on $\, S_{F} \, $ 
(schematically).}
\label{Segments}
\end{figure}

  The lines of directions of $\, {\bf B} \, $, corresponding to 
the emergence of unstable periodic trajectories on $\, S_{F} \, $, 
in particular, are everywhere closely adjacent to the boundaries 
of the Stability Zones $\, \Omega_{\alpha} \, $ (Fig. \ref{Segments}), 
which, in turn, leads to a rather complex behavior of the conductivity 
near the boundaries of $\, \Omega_{\alpha} \, $ (see, e.g., \cite{AnProp}). 
In our consideration here, however, the picture of these lines 
will not play any role, so we will not consider them here. 
The emergence of chaotic trajectories on the Fermi surface is 
unambiguously connected with the structure of the 
``exact mathematical'' Zones $\, \Omega_{\alpha} \, $ on the 
angular diagram (\cite{SecBound,UltraCompl}), which we will 
consider here.

\vspace{1mm}

 Each Stability Zone $\, \Omega_{\alpha} \, $ is obviously a subdomain 
of some Zone $\, W_{\alpha} \, $, such that 
$$ \forall \, {\bf n} \, \in \, \Omega_{\alpha} \,\, : \quad \quad
\widetilde{\epsilon}_{1} ({\bf n}) \,\,\, \leq \,\,\, \epsilon_{F}
\,\,\, \leq \,\,\, \widetilde{\epsilon}_{2} ({\bf n}) $$
(if this set is not empty).

 In general, the Fermi surface diagram is a ``subdiagram'' of 
the full angular diagram, ``taking'' from it what relates to 
the level $\, \epsilon_{F} \, $. By varying the values of 
$\, \epsilon_{F} \, $ in the interval 
$\, \left[ \epsilon_{\rm min} , \, \epsilon_{\rm max} \right] \, $, 
where
$$\epsilon_{\rm min} \,\,\, \leq \,\,\, \epsilon ({\bf p}) 
\,\,\, \leq \,\,\, \epsilon_{\rm max}  \,\,\, , $$
we obtain a set of angular diagrams for the Fermi surfaces 
depending on the parameter $\, \epsilon_{F} \, $.

 In the generic case, the angular diagrams of the Fermi surface 
are very simple for values of $\, \epsilon_{F} \, $ close to 
$\, \epsilon_{\rm min} \, $ or $\, \epsilon_{\rm max} \, $. 
Namely, since the Fermi surfaces are in this case (small) spheres, 
they do not contain open trajectories of (\ref{MFSyst}), and the 
entire angular diagram shows the presence of only closed 
trajectories on $\, S_{F} \, $. Here we will be interested only in 
sufficiently complex angular diagrams corresponding to sufficiently 
complex Fermi surfaces (Fig. \ref{ArbSurf}). It is easy to see that
the values of $\, \epsilon_{F} \, $ should in this case be somewhat
distant from the boundaries of the interval
$\, \left[ \epsilon_{\rm min} , \, \epsilon_{\rm max} \right] \, $.

 In our sense, the angular diagram of a Fermi surface is 
sufficiently complex if it contains a non-empty set of Stability 
Zones $\, \Omega_{\alpha} \, $. It is easy to see that the Fermi 
energy must then satisfy the condition
$$\epsilon^{\cal A}_{1} \,\,\, < \,\,\, \epsilon_{F} 
\,\,\, < \,\,\, \epsilon^{\cal A}_{2} \,\,\, , $$
where
$$\epsilon^{\cal A}_{1} \,\,\, = \,\,\, \min \,\,
\widetilde{\epsilon}_{1}({\bf n}) \,\,\, , \quad
\epsilon^{\cal A}_{2} \,\,\, = \,\,\, \max \,\,
\widetilde{\epsilon}_{2}({\bf n}) $$

 We will, however, be interested in angular diagrams 
containing not only the Zones $\, \Omega_{\alpha} \, $, 
but also directions of $\, {\bf B} \, $, corresponding to 
the emergence of chaotic trajectories on the Fermi surface. 
As it turns out (see \cite{SecBound,UltraCompl}), for generic 
dispersion relations (with an infinite number of Zones 
$\, W_{\alpha} $) such diagrams arise in an even narrower interval
$$\epsilon_{F} \,\,\, \in \,\,\, \left[ \epsilon^{\cal B}_{1} , \,
\epsilon^{\cal B}_{2} \right] \,\,\, , $$
where 
\begin{equation}
\label{CalBrel}
\epsilon^{\cal B}_{1} \,\,\, = \,\,\, \min \,\,
\widetilde{\epsilon}_{2}({\bf n}) \,\,\, , \quad
\epsilon^{\cal B}_{2} \,\,\, = \,\,\, \max \,\,
\widetilde{\epsilon}_{1}({\bf n})  \,\,\, , \quad 
\end{equation}
$$\epsilon^{\cal A}_{1} \,\,\, < \,\,\,
\epsilon^{\cal B}_{1} \,\,\, < \,\,\,
\epsilon^{\cal B}_{2} \,\,\, < \,\,\, 
\epsilon^{\cal A}_{2}  \quad \quad \quad  $$

 Moreover, all sufficiently complex angular diagrams of 
Fermi surfaces can be divided into two classes.

\vspace{1mm}

\noindent
1) Type A diagrams:
$$\epsilon_{F} \,\,\, \in \,\,\, 
\left( \epsilon^{\cal A}_{1} , \, \epsilon^{\cal B}_{1} \right) 
\cup \left( \epsilon^{\cal B}_{2} , \, \epsilon^{\cal A}_{2} \right) $$

\begin{figure}[t]
\begin{center}
\includegraphics[width=\linewidth]{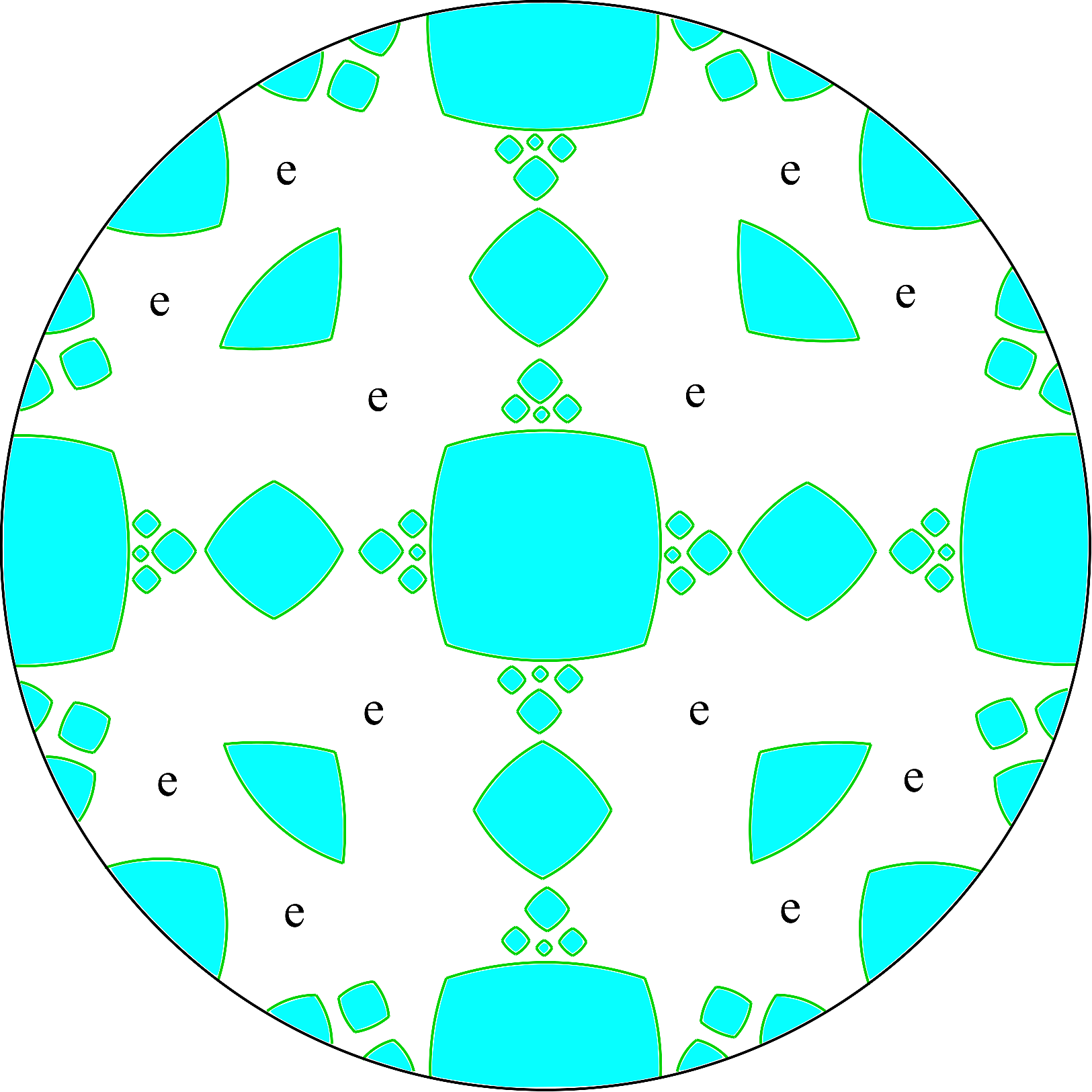}
\end{center}
\begin{center}
\includegraphics[width=\linewidth]{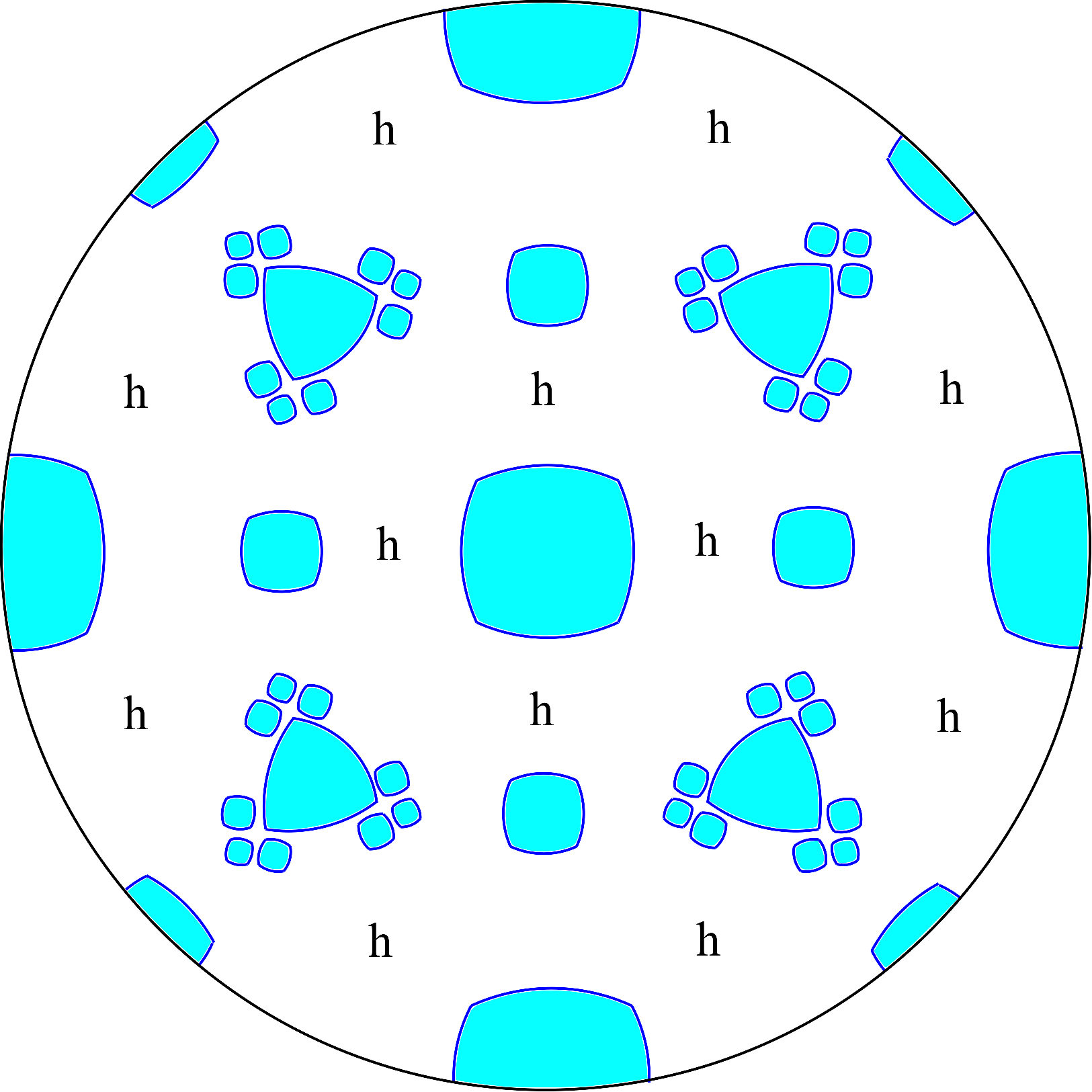}
\end{center}
\caption{Diagrams of type A for
$\, \epsilon_{F} \, \in \, 
\left( \epsilon^{\cal A}_{1} , \, \epsilon^{\cal B}_{1} \right) \, $
and $\, \epsilon_{F} \, \in \, 
\left( \epsilon^{\cal B}_{2} , \, \epsilon^{\cal A}_{2} \right)$ 
(schematically). The signs e and h indicate the types of Hall 
conductivity for directions of $\, {\bf B} \, $, corresponding 
to the presence of only closed trajectories on the Fermi surface.
}
\label{DiagramsA}
\end{figure}

 Diagrams of type A contain a finite number of Stability Zones 
$\, \Omega_{\alpha} \, $ and do not contain directions of 
$\, {\bf B} \, $ corresponding to the emergence of chaotic 
trajectories on the Fermi surface. For all directions of 
$\, {\bf B} \, $ (of maximal irrationality) corresponding to 
the presence of only closed trajectories on the Fermi surface, 
the Hall conductivity $\, \sigma^{xy} ({\bf B}) \, $ has 
the same type at $\, B \rightarrow \infty \, $ (electron type for 
$\, \epsilon_{F} \, \in \,
\left( \epsilon^{\cal A}_{1} , \, \epsilon^{\cal B}_{1} \right) \, $ 
and hole type for 
$\, \epsilon_{F} \, \in \, 
\left( \epsilon^{\cal B}_{2} , \, \epsilon^{\cal A}_{2} \right)$) 
(Fig. \ref{DiagramsA}). It is natural to call such diagrams 
diagrams of type ${\rm A}_{-}$ for 
$\, \epsilon_{F} \, \in \, 
\left( \epsilon^{\cal A}_{1} , \, \epsilon^{\cal B}_{1} \right) \, $ 
and diagrams of type ${\rm A}_{+}$ for 
$\, \epsilon_{F} \, \in \, 
\left( \epsilon^{\cal B}_{2} , \, \epsilon^{\cal A}_{2} \right)$.

\vspace{1mm}

\noindent
1) Type B diagrams:
$$\epsilon_{F} \,\,\, \in \,\,\, 
\left[ \epsilon^{\cal B}_{1} , \, \epsilon^{\cal B}_{2} \right] $$

\begin{figure}[t]
\begin{center}
\includegraphics[width=\linewidth]{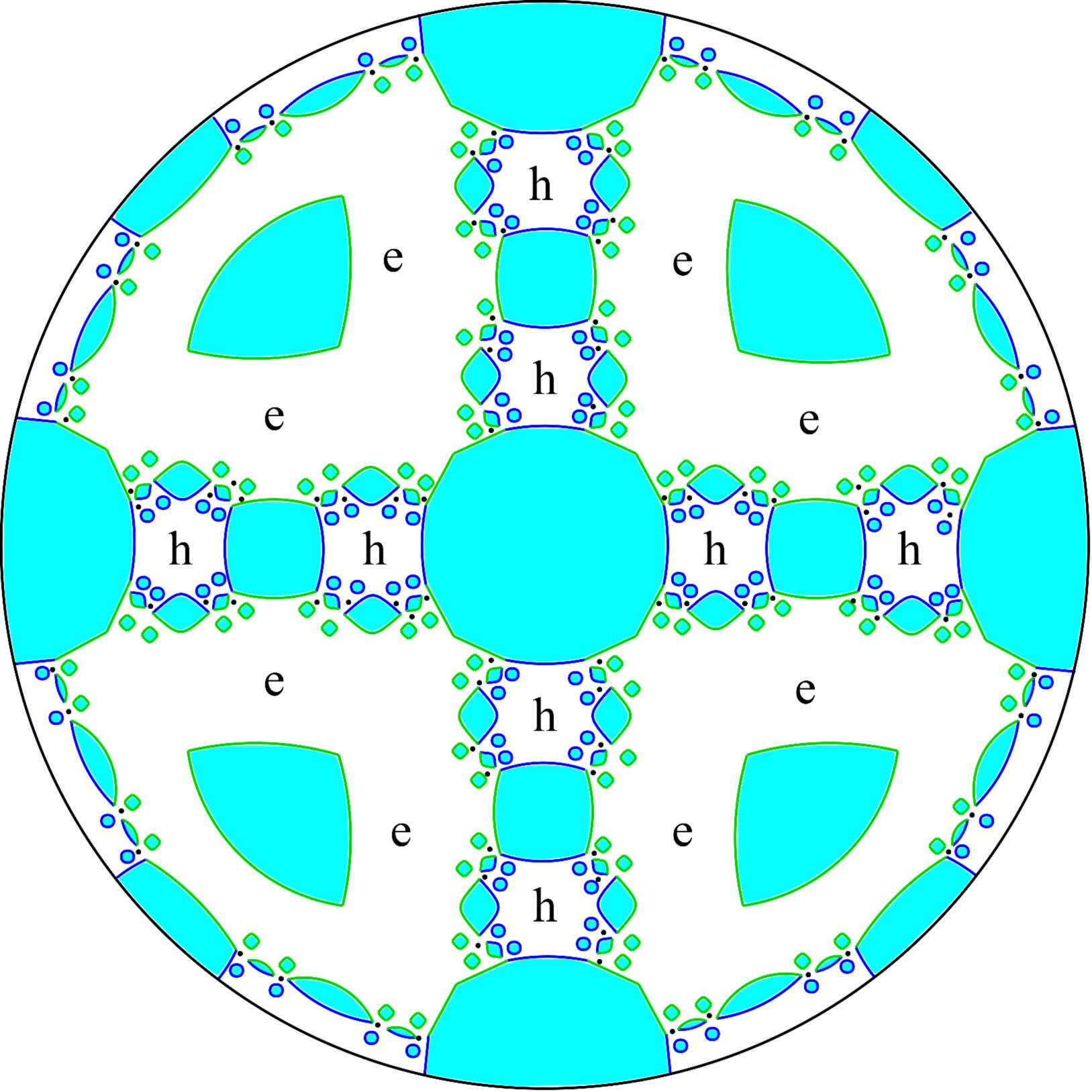}
\end{center}
\begin{center}
\includegraphics[width=\linewidth]{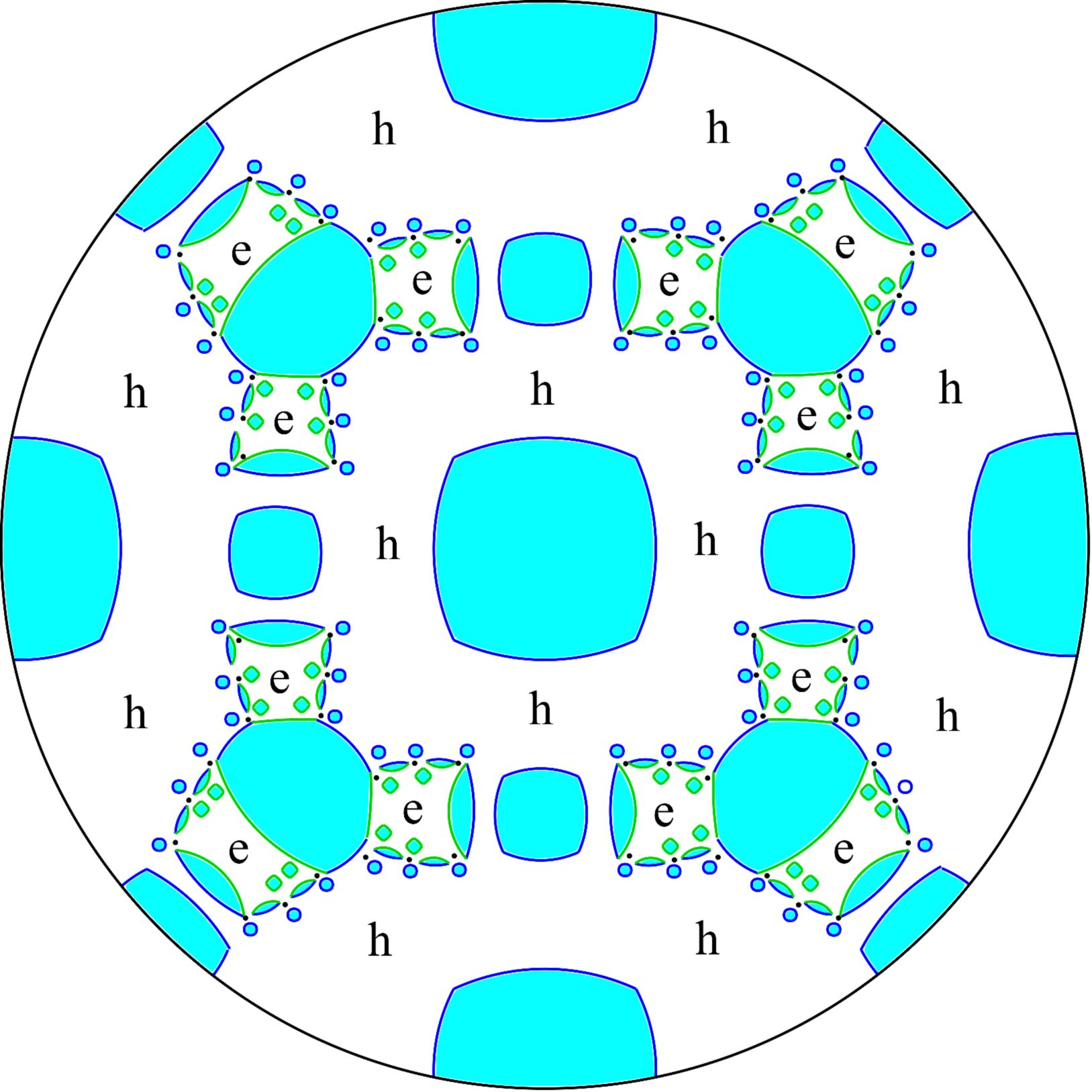}
\end{center}
\caption{Diagrams of type B
($\epsilon_{F} \, \in \, 
\left( \epsilon^{\cal B}_{1} , \, \epsilon^{\cal B}_{2} \right)$)
(schematically). The signs e and h indicate the types of Hall 
conductivity for directions of $\, {\bf B} \, $, corresponding 
to the presence of only closed trajectories on the Fermi surface.
}
\label{DiagramsB}
\end{figure}

 Type B diagrams contain an infinite number of Stability Zones 
$\, \Omega_{\alpha} \, $, as well as directions of $\, {\bf B} \, $, 
corresponding to the emergence of chaotic trajectories on the Fermi 
surface. Among the regions on $\, \mathbb{S}^{2} \, $, corresponding 
to the presence of only closed trajectories on the Fermi surface, 
there are regions of both electron and hole Hall conductivity 
(except for the values $\, \epsilon_{F} = \epsilon^{\cal B}_{1} \, $
and $\, \epsilon_{F} = \epsilon^{\cal B}_{2} $) (Fig. \ref{DiagramsB}).

\vspace{1mm}

 As can be seen, the Fermi surfaces with B-type diagrams cannot be 
definitely assigned to the electron or hole type 
(this property depends here on the direction of $\, {\bf B} $). 
When $\, \epsilon_{F} \, $ varies in the interval 
$\, \left( \epsilon^{\cal B}_{1} , \, \epsilon^{\cal B}_{2} \right) \, $ 
the shape of the B-type diagram changes very quickly, providing 
a transition from the ${\rm A}_{-}$ diagram to the ${\rm A}_{+}$ 
diagram (Fig. \ref{DiagramsB}).

 Type B diagrams are interesting from many points of view, 
particularly in relation to the observation of chaotic conductivity 
behavior in the limit $\, \omega_{B} \tau \rightarrow \infty \, $. 
However, such diagrams have not yet been observed experimentally. 
In our view, this may be due to the small width of the interval 
$\, \left[ \epsilon^{\cal B}_{1} , \, \epsilon^{\cal B}_{2} \right] \, $ 
for real dispersion relations. Investigating the position of the interval 
$\, \left[ \epsilon^{\cal B}_{1} , \, \epsilon^{\cal B}_{2} \right] \, $ 
for dispersion relations of various types can provide crucial information 
for searching such diagrams and the conductivity behavior 
described above.

\vspace{1mm}

 This paper investigates the probability of chaotic conductivity 
behavior in the tight-binding approximation (see, for example, 
\cite{Kittel,Ziman,AshcroftMermin,etm}) for crystals with cubic 
symmetry. More precisely, we consider the tight-binding approximation 
for simple, body-centered, and face-centered cubic lattices.

 The tight-binding approximation for a simple cubic lattice 
(see \cite{Kittel, Ziman}) gives, in leading order, the following 
expression for the dispersion relation
$$\epsilon({\bf p}) \,\,\,\, = \,\,\,\, \cos {p_{x} l \over \hbar} \,\, + \,\,
\cos {p_{y} l \over \hbar} \,\, + \,\, \cos {p_{z} l \over \hbar} $$

(Since we are interested here only in the ratio of the width of the interval 
$\, \left[ \epsilon^{\cal B}_{1} , \, \epsilon^{\cal B}_{2} \right] \, $
to the size of the interval
$\, \left[ \epsilon_{\max} , \, \epsilon_{\min} \right] \, $,
we have omitted the constant term in the above expression and also 
normalized the coefficient of the main remaining term to 1).

 The dispersion relation given above is a non-generic relation 
and corresponds to the situation 
$\, \epsilon^{\cal B}_{1} = \epsilon^{\cal B}_{2} \, $. This applies, 
in fact, to all dispersion relations containing only odd harmonics 
($k_{1} + k_{2} + k_{3} \, = \, 2 m + 1$) in the Fourier expansion. 
The reason is that the shift
$$p_{x} \,\, \rightarrow \,\, p_{x} \, + \, {\pi \hbar \over l} \,\, , \quad
p_{y} \,\, \rightarrow \,\, p_{y} \, + \, {\pi \hbar \over l} \,\, , \quad
p_{z} \,\, \rightarrow \,\, p_{z} \, + \, {\pi \hbar \over l} $$
is equivalent in this case to the replacement
$$\epsilon({\bf p}) \,\,\, \rightarrow \,\,\, - \, \epsilon({\bf p})
\,\,\, , $$
which, in turn, entails the relation
$$\epsilon_{1} ({\bf n}) \,\,\, = \,\,\, - \, \epsilon_{2} ({\bf n}) $$
for any direction of $\, {\bf B} \, $. As a consequence, the relation 
$\, \epsilon_{0} ({\bf n}) = 0 \, $ also arises for all the ``chaotic'' 
directions. It can be seen, therefore, that in the tight-binding 
approximation for a simple cubic lattice, the width of the interval 
$\, \left[ \epsilon^{\cal B}_{1} , \, \epsilon^{\cal B}_{2} \right] \, $ 
in the leading order vanishes, which corresponds to zero probability 
of $\, \epsilon_{F} \, $ falling within this interval.

 The angular diagram for the Fermi surface $\, \epsilon_{F} = 0 \, $ 
coincides in this case with the full angular diagram (for the entire 
dispersion relation) and is the most complex (the results of its 
numerical study are presented, for example, 
in \cite{DynnBuDA,DeLeo2017,DynMalNovUMN}).

 All of the above also applies to the body-centered cubic lattice, 
where the tight-binding approximation (see \cite{Kittel}) leads in 
the leading order to the relation
$$\epsilon({\bf p}) \,\,\,\, = \,\,\,\, \cos {p_{x} l \over \hbar} \,\, 
\cos {p_{y} l \over \hbar} \,\, \cos {p_{z} l \over \hbar} $$

 It can be seen, therefore, that for both the simple and 
body-centered cubic lattices the nonzero length of the interval 
$\, \left[ \epsilon^{\cal B}_{1} , \, \epsilon^{\cal B}_{2} \right] \, $ 
is due to higher-order corrections and is very small. The interval 
$\, \left[ \epsilon^{\cal B}_{1} , \, \epsilon^{\cal B}_{2} \right] \, $ 
is located near the zero energy value (at the center of the conduction band). 
Observation of chaotic regimes in such conductors may probably be possible 
by artificially shifting the Fermi energy (for example, by applying external 
pressure) to the desired value.

\vspace{1mm}

 Below we present an estimate of the width of the interval 
$\, \left[ \epsilon^{\cal B}_{1} , \, \epsilon^{\cal B}_{2} \right] \, $
for the dispersion relation
\begin{multline}
\label{DispRel}
\epsilon({\bf p}) \,\,\,\, = \,\,\,\, 
\cos {p_{x} l \over \hbar} \, \cos {p_{y} l \over \hbar}
\,\, +  \,\, \cos {p_{y} l \over \hbar} \, \cos {p_{z} l \over \hbar} 
\,\, +  \\
+ \,\, \cos {p_{x} l \over \hbar} \, \cos {p_{z} l \over \hbar} 
\,\,\, , 
\end{multline}
which is a tight-binding approximation (see e.g. \cite{Kittel,AshcroftMermin}) 
for crystals with a face-centered cubic cell.

 As is not difficult to see, in this case we have
$$\epsilon_{\min} \,\, = \,\, - 1 \,\,\, , \quad
\epsilon_{\max} \,\, = \,\, 3 $$

 The relation (\ref{DispRel}) corresponds to a nonzero 
width of the interval 
$\, \left[ \epsilon^{\cal B}_{1} , \, \epsilon^{\cal B}_{2} \right]$. 
At the same time, the function (\ref{DispRel}) is not a Morse function 
at the level $\, \epsilon({\bf p}) = - 1 \, $, so crossing the 
value $\, -1 \, $ immediately leads to the emergence of surfaces
$$\epsilon({\bf p}) \,\,\, = \,\,\, - 1 \,\, + \,\, \delta \,\,\, ,$$
extending in all directions in $\, {\bf p}$ - space. This situation 
persists up to the value $\, \epsilon_{F} = 0 \, $, 
for $\, \epsilon_{F} > 0 \, $ the Fermi surfaces are spheres 
in $\, {\bf p}$ - space, unconnected to each other. Open trajectories 
of system (\ref{MFSyst}) arise here in the entire interval 
$\, (-1, 0) \, $, so we have
$$\epsilon^{\cal A}_{1} \,\, = \,\, - 1 \,\,\, , \quad
\epsilon^{\cal A}_{2} \,\, = \,\, 0 $$

 The complete angular diagram of the dispersion relation 
(\ref{DispRel}) contains an infinite number of Stability Zones 
$\, W_{\alpha} \, $, its rather complex form (obtained as a result of 
serious numerical studies) is presented in the work \cite{DeLeo2017}.

\vspace{1mm}

 According to (\ref{CalBrel}), to find the interval 
$\, \left[ \epsilon^{\cal B}_{1} , \, \epsilon^{\cal B}_{2} \right] \, $, 
we need to know the values of $\, \widetilde{\epsilon}_{1} ({\bf n}) \, $
and $\, \widetilde{\epsilon}_{2} ({\bf n}) \, $ everywhere on the unit 
sphere $\, \mathbb{S}^{2} \, $. Their calculation on $\, \mathbb{S}^{2} \, $, 
however, requires very serious numerical studies, and we will use here 
a certain technique that allows us to give an estimate close to the 
exact one (and possibly coinciding with it) 
for the position of the interval 
$\, \left[ \epsilon^{\cal B}_{1} , \, \epsilon^{\cal B}_{2} \right] \, $ 
for the given dispersion relation.

 First of all, for many dispersion relations, the values 
$\, \min \, \widetilde{\epsilon}_{2} ({\bf n}) \, $ and 
$\, \max \, \widetilde{\epsilon}_{1} ({\bf n}) \, $ can be 
calculated not over the entire sphere $\, \mathbb{S}^{2} \, $, 
but only over the boundaries of the zones $\, W_{\alpha} \, $. 
We note immediately that everywhere on the boundaries of
$\, W_{\alpha} \, $ 
$$\widetilde{\epsilon}_{1} ({\bf n}) \,\,\, = \,\,\, 
\widetilde{\epsilon}_{2} ({\bf n}) \,\,\, \equiv \,\,\, 
\widetilde{\epsilon}_{0} ({\bf n}) \,\,\, , $$
so we have the relation
$$\epsilon^{\cal B}_{1} \,\,\, < \,\,\,
\epsilon^{\cal B}_{2}  $$
for any non-constant function
$\, \widetilde{\epsilon}_{0} ({\bf n}) \, $ on
$\, \partial W_{\alpha} \, $. 

 If for any of the Zones $\, W_{\alpha} \, $
$$\min_{W_{\alpha}} \, \widetilde{\epsilon}_{2} ({\bf n}) 
\quad < \quad 
\min_{\partial W_{\alpha}} \, \widetilde{\epsilon}_{2} ({\bf n}) 
\quad = \quad
\min_{\partial W_{\alpha}} \, \widetilde{\epsilon}_{1} ({\bf n}) $$
or 
$$\max_{W_{\alpha}} \, \widetilde{\epsilon}_{1} ({\bf n}) 
\quad > \quad 
\max_{\partial W_{\alpha}} \, \widetilde{\epsilon}_{1} ({\bf n}) 
\quad = \quad
\max_{\partial W_{\alpha}} \, \widetilde{\epsilon}_{2} ({\bf n}) 
\,\,\, , $$
we should observe the emergence of non-simply connected zones 
$\, \Omega_{\alpha} \, $ for some values of $\, \epsilon_{F} \, $. 
Such a situation is possible for sufficiently complex 
(specially constructed) dispersion relations 
$\, \epsilon ({\bf p}) \, $ (see, for example, \cite{DynMalNovUMN}), 
however, it is extremely unlikely for real dispersion laws in crystals.

 The value
\begin{equation}
\label{Wdiff}
\max_{\partial W_{\alpha}} \, \widetilde{\epsilon}_{1} ({\bf n}) 
\, - \, 
\min_{\partial W_{\alpha}} \, \widetilde{\epsilon}_{2} ({\bf n})
\quad = \quad 
\max_{\partial W_{\alpha}} \, \widetilde{\epsilon}_{0} ({\bf n}) 
\, - \, 
\min_{\partial W_{\alpha}} \, \widetilde{\epsilon}_{0} ({\bf n}) 
\end{equation}
is determined by the variation of the function 
$\, \widetilde{\epsilon}_{0} ({\bf n}) \, $ on the boundary of 
$\, W_{\alpha} \, $ and, in general, is larger for large zones 
$\, W_{\alpha} \, $. Calculating the value (\ref{Wdiff}) for the 
largest zones $\, W_{\alpha} \, $ gives a good (and often accurate) 
estimate for the width of the interval
$\, \left[ \epsilon^{\cal B}_{1} , \, \epsilon^{\cal B}_{2} \right] \, $.

\begin{figure}[t]
\begin{center}
\includegraphics[width=\linewidth]{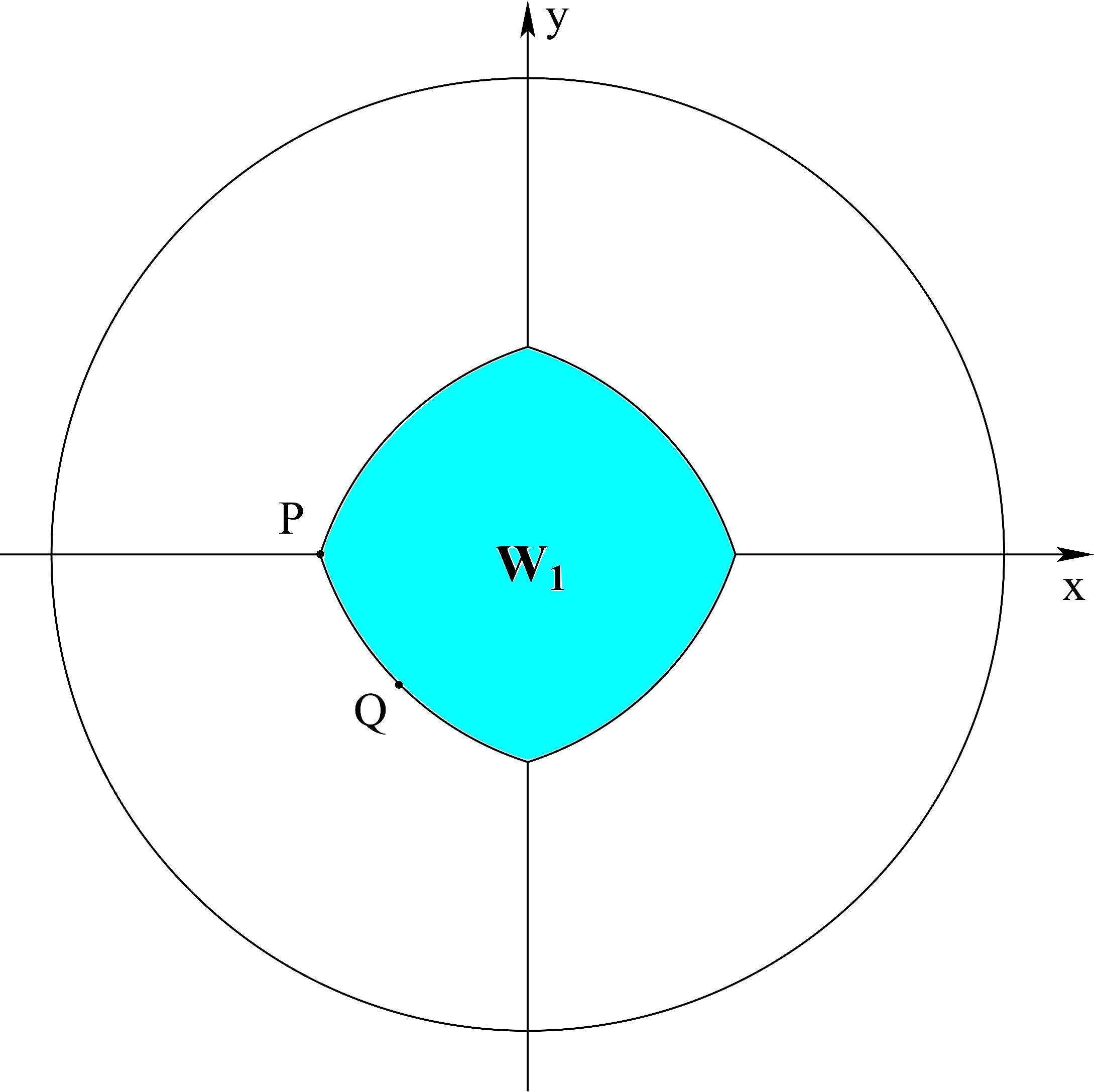}
\end{center}
\caption{Zone $\, W_{1} \, $ on the angular diagram of the 
dispersion relation (\ref{DispRel}).}
\label{W1}
\end{figure}

 Here we will calculate the values of 
$\, \widetilde{\epsilon}_{0} ({\bf n}) \, $ at the most distant 
boundary points of the Stability Zone $\, W_{1} \, $, which inherits 
the symmetry of the dispersion relation (\ref{DispRel}) and is the 
largest of the Zones $\, W_{\alpha} \, $. We also note here that 
the existence of such Stability Zones for dispersion relations 
possessing rotational symmetry was indicated in the work (\cite{dynn3}) 
and is a fairly general fact. The Zone that we will consider contains 
the point $\, (0, 0, 1) \in \mathbb{S}^{2} \, $ and is ``square'' in 
shape (Fig. \ref{W1}).

 We use here the values of $\, \widetilde{\epsilon}_{0} ({\bf n}) \, $
at the most symmetric points of the boundary of $\, W_{1} \, $ 
($P \, $ and $\, Q$), where their calculation is somewhat simplified. 
As we have already said, for the interval 
$\, \left[ \epsilon^{\cal B}_{1} , \,\epsilon^{\cal B}_{2} \right] \, $ 
we will use the estimate
$\, \left[ \widetilde{\epsilon}_{0} (P) , \, 
\widetilde{\epsilon}_{0} (Q) \right] \, $. The properties of the functions 
$\, \widetilde{\epsilon}_{1} ({\bf n}) \, $ and 
$\, \widetilde{\epsilon}_{2} ({\bf n}) \, $ allow us to expect that 
such an estimate is quite good for dispersion relations that have a high 
symmetry (and for many relations, in fact, exact).

\vspace{1mm}

\begin{figure}[t]
\begin{center}
\includegraphics[width=\linewidth]{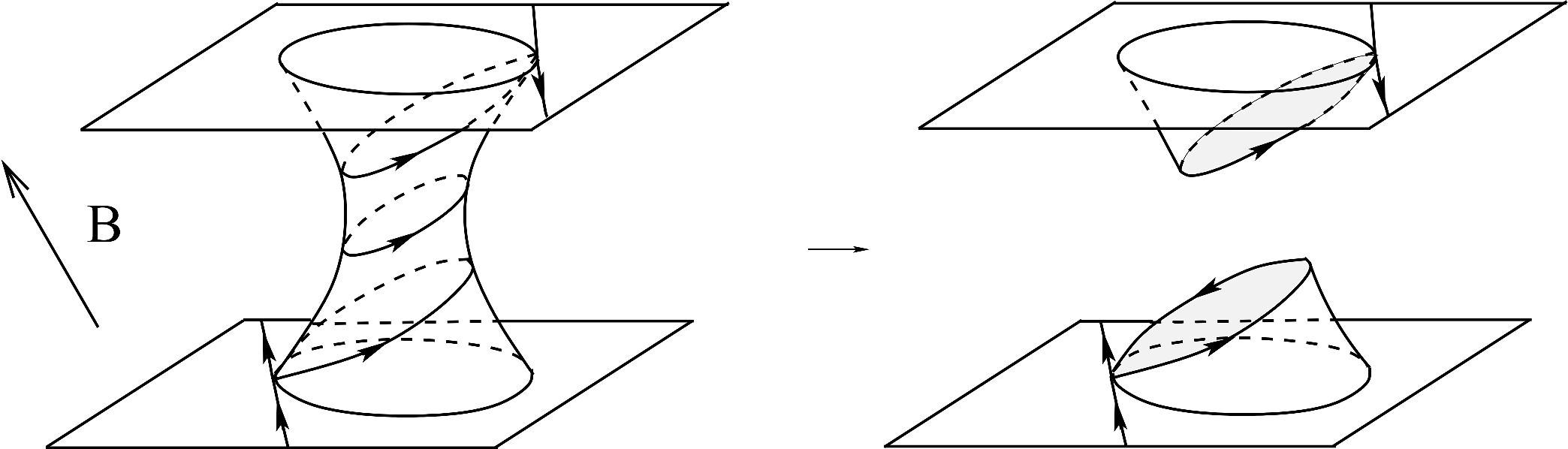}
\end{center}
\caption{Fermi surface reduction procedure (schematically).
}
\label{Reduction}
\end{figure}

 To study the trajectories of system (\ref{MFSyst}), we use the Fermi 
surface reduction method (\cite{dynn3}). Specifically, for each 
direction of $\, {\bf B} \, $, we remove cylinders of closed trajectories 
$\, \left\{ C_{i} \right\} \, $ from the Fermi surface and then seal the 
resulting holes with flat disks orthogonal to $\, {\bf B} \, $ 
(Fig. \ref{Reduction}). The new ``reduced'' Fermi surface carries the 
open trajectories of system (\ref{MFSyst}), and its study allows us to 
determine both the type of open trajectories and their other characteristics.

 It is convenient also to consider system (\ref{MFSyst}) in the torus 
$ \, \mathbb{T}^{3} \, = \, \mathbb{R}^{3} \big/ L^{*} \, $. 
The set of cylinders $\, \left\{ C_{i} \right\} \, $ then becomes finite, 
and the Fermi surface (as well as the reduced Fermi surface) becomes 
a compact two-dimensional surface embedded in $\, \mathbb{T}^{3} \, $.

\begin{figure}[t]
\begin{center}
\includegraphics[width=\linewidth]{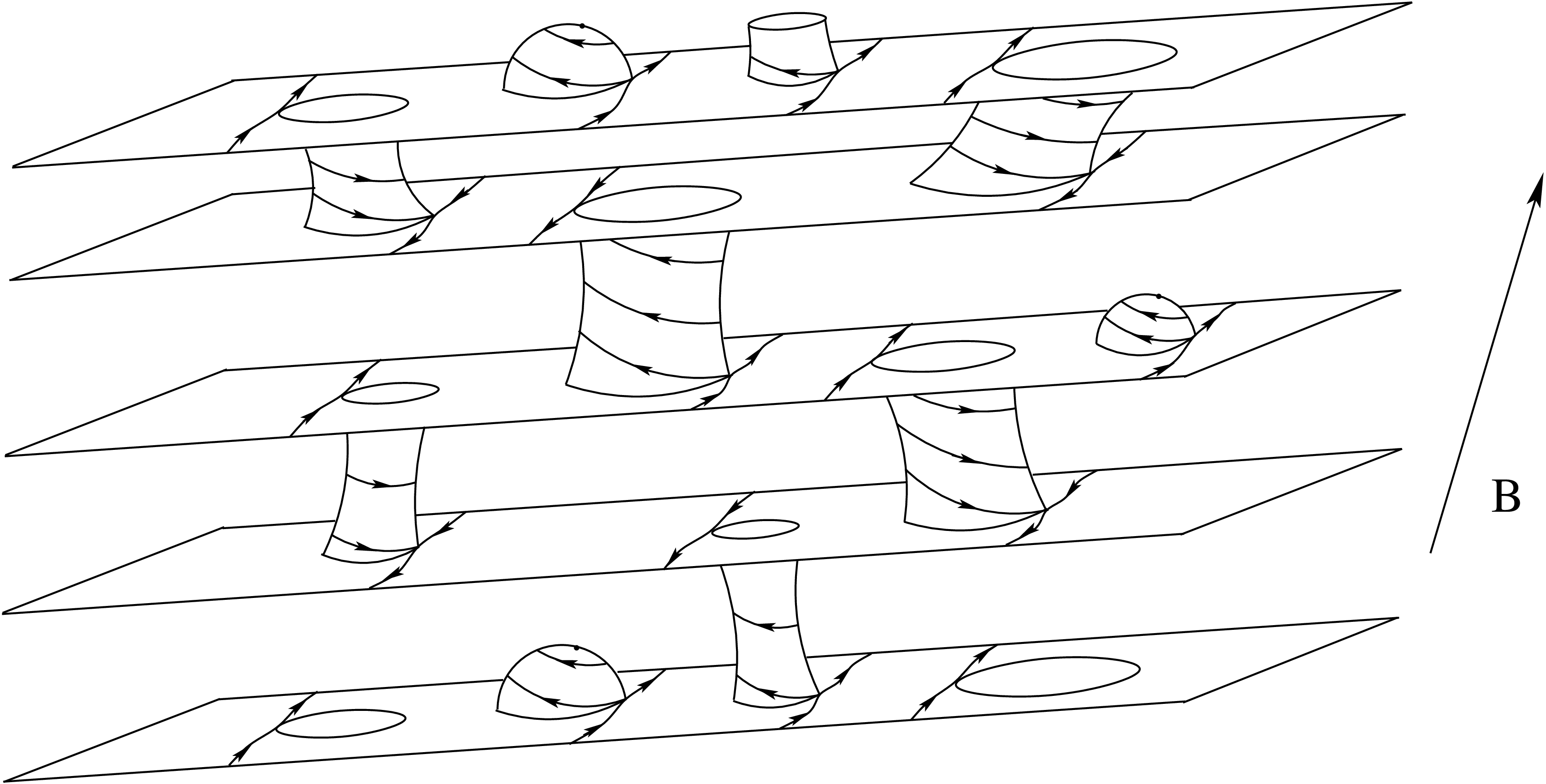}
\end{center}
\caption{Cutting the Fermi surface into carriers of stable open 
trajectories by cylinders $\, C_{i} \, $ for direction $\, {\bf B} \, $ 
lying in one of the Stability Zones $\, \Omega_{\alpha} \, $ (schematically).
}
\label{FullFermSurf}
\end{figure}

 The most important characteristic of the reduced Fermi surface is 
the genus $\, g \, $ of its connected components. In particular, 
the decomposition of the reduced Fermi surface into connected 
components of genus $\, g = 1 \, $ (two-dimensional tori 
$\, \mathbb{T}^{2}$) corresponds to the presence of stable open 
trajectories of (\ref{MFSyst}) (Fig. \ref{FullFermSurf}). The case 
$\, g \geq 3 \, $ corresponds to the emergence of chaotic trajectories
on the Fermi surface (the disappearance of the Fermi surface during 
the reduction process corresponds to the presence of only closed 
trajectories of the system (\ref{MFSyst})). 

\begin{figure}[t]
\begin{center}
\includegraphics[width=\linewidth]{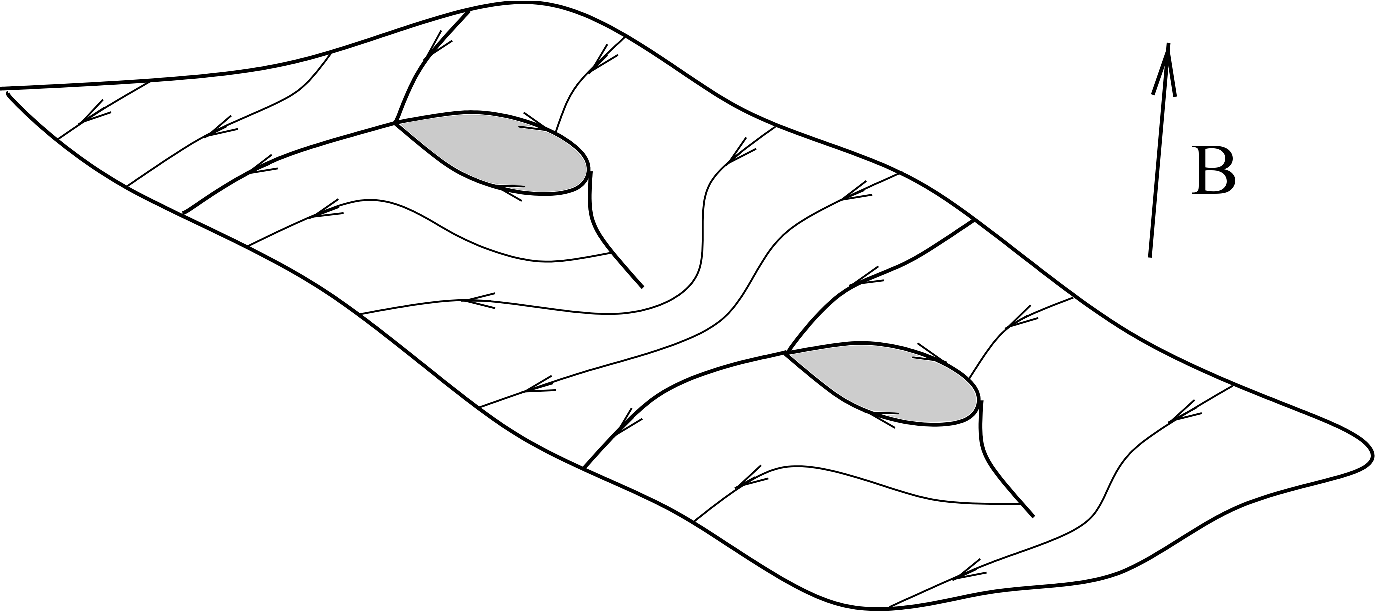}
\end{center}
\caption{The shape of a carrier of stable open trajectories of 
system (\ref{MFSyst}) in the $\, {\bf p}$ - space (schematically).
}
\label{IntPlane}
\end{figure}

 We also note here that 
the carriers of stable open trajectories (two-dimensional tori 
$\, \mathbb{T}^{2}$) have the form of periodically deformed planes 
in the full $\, {\bf p}$ - space, having a common integer 
(generated by two reciprocal lattice vectors) two-dimensional direction 
(Fig. \ref{IntPlane}). The total number of such tori in 
$\, \mathbb{T}^{3} \, $ is even for any physical Fermi surface.

 A change in the set of cylinders $\, \left\{ C_{i} \right\} \, $ 
under rotations of $\, {\bf B} \, $ can be called a change in the 
topological structure of system (\ref{MFSyst}) on the Fermi surface. 
In particular, the leaving a Stability Zone $\, \Omega_{\alpha} \, $ 
(to the region of the presence of only closed trajectories on 
$\, S_{F}$) is always determined by the disappearance of one of the 
cylinders $\, C_{i} \, $ from the set $\, \left\{ C_{i} \right\} \, $ 
(and the emergence of new cylinders).

\vspace{1cm}

 As for the boundaries of the Stability Zones $\, W_{\alpha} \, $ 
for the whole dispersion relation $\, \epsilon ({\bf p}) \, $, 
they are determined by the disappearance of at least two cylinders 
$\, C_{i} \, $ on the surface 
$\, \epsilon ({\bf p}) \, = \widetilde{\epsilon}_{0} ({\bf n}) \, $ 
(or even more in the presence of additional symmetries). 
The vanishing of the height of two (or more) cylinders 
$\, C_{i} \, $ requires a special selection of the value 
$\, \widetilde{\epsilon}_{0} ({\bf n}) \, $ 
(together with the direction of $\, {\bf B}$), 
which determines the functions 
$\, \widetilde{\epsilon}_{1} ({\bf n}) = 
\widetilde{\epsilon}_{2} ({\bf n}) = 
\widetilde{\epsilon}_{0} ({\bf n}) \, $ 
on the boundary of $\, W_{\alpha} \, $.

\vspace{1cm}

\begin{figure}[t]
\begin{center}
\includegraphics[width=0.7\linewidth]{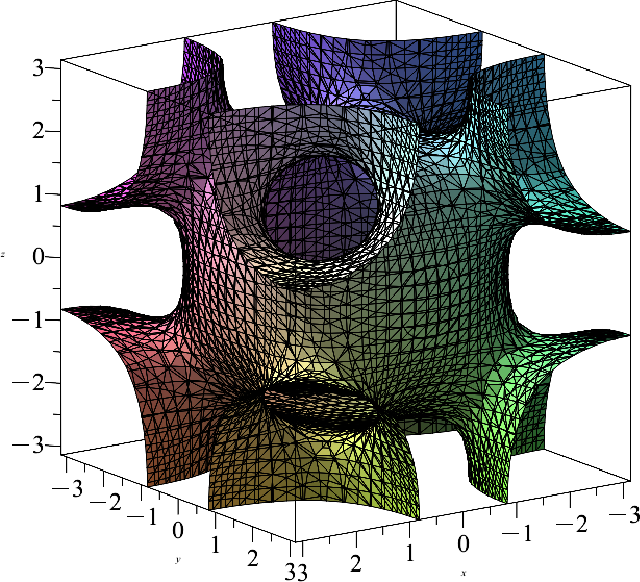}
\end{center}
\caption{The Fermi surface $\, \cos x \, \cos y \,\, + \,\,
\cos y \, \cos z \,\, + \,\,  \cos x \, \cos z \,\, = \,\, 
- 0.35 \, $.
}
\label{Surface}
\end{figure}

 All Fermi surfaces in the interval 
$\, \left( \epsilon^{\cal A}_{1} , \, \epsilon^{\cal A}_{2} \right) \, $ 
have genus 4 (Fig. \ref{Surface}). This means, in particular, that for any 
$\, {\bf n} \in W_{\alpha} \, $ and 
$\, \epsilon_{F} \in \left[ \tilde{\epsilon}_{1} ({\bf n}) , \,
\tilde{\epsilon}_{2} ({\bf n}) \right] \, $ 
the Fermi surface
$$\epsilon ({\bf p}) \,\,\, = \,\,\, \epsilon_{F} $$
splits into two (non-equivalent) carriers of open trajectories 
$\, \left\{ \mathbb{T}^{2}_{1} , \, \mathbb{T}^{2}_{2} \right\} \, $ 
and three cylinders of closed trajectories 
$\, \left\{ C_{1} , \, C_{2} , \, C_{3} \right\} \, $. 
As we have already said, the boundaries of the Zones $\, W_{\alpha} \, $ 
correspond to the vanishing of the heights of at least two cylinders from 
the set $\, \left\{ C_{1} , \, C_{2} , \, C_{3} \right\} \, $ at the level 
$\, \widetilde{\epsilon}_{1} ({\bf n}) = 
\widetilde{\epsilon}_{2} ({\bf n}) = 
\widetilde{\epsilon}_{0} ({\bf n}) \, $.

\begin{figure}[t]
\begin{center}
\includegraphics[width=0.8\linewidth]{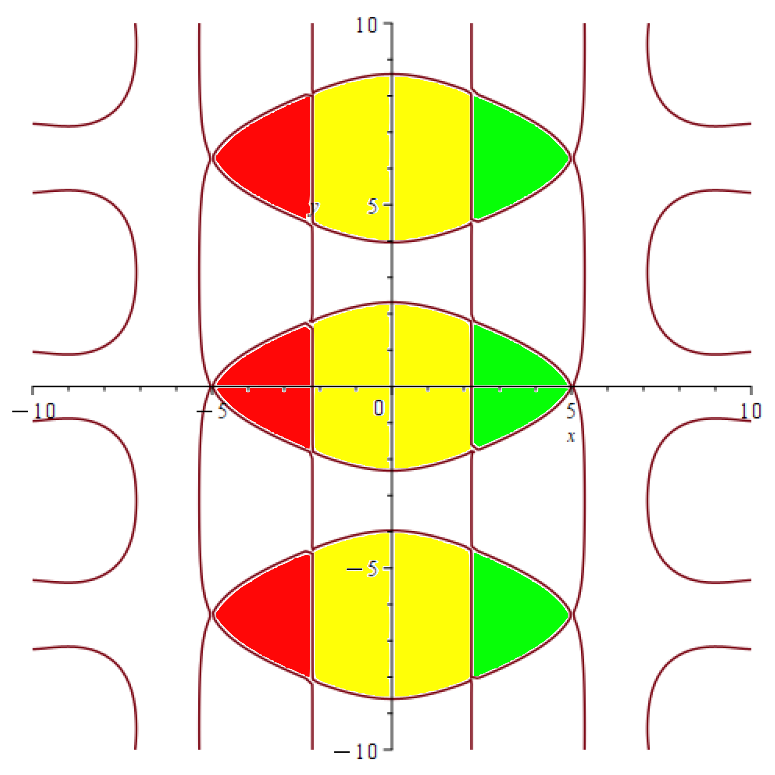}
\end{center}
\caption{Level lines (\ref{PointPLines}) in the plane passing through 
the origin, for
$\, {\bf n} = P \, $ and
$\, \epsilon_{F} \, = \, \widetilde{\epsilon}_{0} (P) 
\, \simeq \, - 0.359 \, $.
}
\label{PointP}
\end{figure}

 At point $P$, due to additional symmetry, the heights of all three 
cylinders $\, C_{1} , \, C_{2} , \, C_{3} \, $ cutting the Fermi 
surface into carriers of open trajectories, vanish at the level 
$\, \widetilde{\epsilon}_{0} (P) \, $. The corresponding cylinders 
of zero height (containing saddle singular points) are present 
(for example) in the plane passing through the origin (Fig. \ref{PointP}). 
The corresponding level lines $\, \epsilon ({\bf p}) \, $ in this plane 
are given by the equation
\begin{equation}
\label{PointPLines}
\cos x \,\, \cos y \,\,\, + \,\,\, (\cos x \, + \,  \cos y) \,\,
\cos 0.419 x \,\,\,  = \,\,\, - 0.359  \,\,\, , 
\end{equation}
which determines both the position of point $P$ and the value
$\, \widetilde{\epsilon}_{0} (P) = - 0.359 $.

\begin{figure}[t]
\begin{center}
\includegraphics[width=0.8\linewidth]{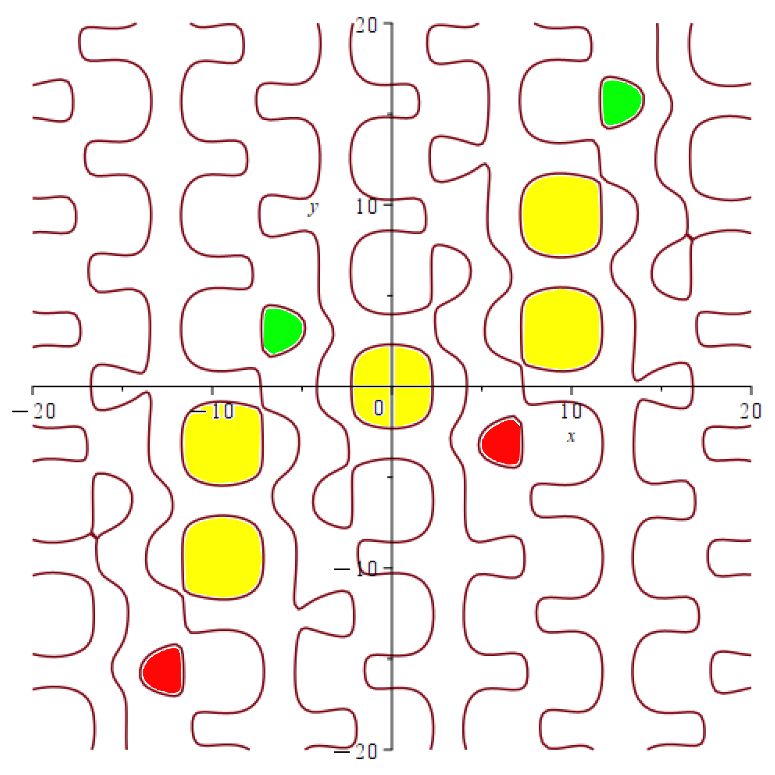}
\end{center}
\caption{The level lines $\, \cos x \, \cos y \,\, + \,\, 
(\cos x + \cos y) \, \cos (0.3 x + 0.05 y)  
\,\, = \,\, - 0.359 $
}
\label{CloseP}
\end{figure}

 For comparison, Fig. \ref{CloseP} shows the level lines of the 
function $\, \epsilon ({\bf p}) \, $ in one of the planes for 
direction $\, {\bf n} \in W_{\alpha} \, $, close to $P$, at the 
same energy level. It can be seen that this plane has nonsingular 
intersections with the cylinders $\, C_{1} , \, C_{2} , \, C_{3} \, $, 
indicating their finite heights inside $\, W_{\alpha} \, $.

\vspace{1mm}

\begin{figure}[t]
\begin{center}
\includegraphics[width=0.8\linewidth]{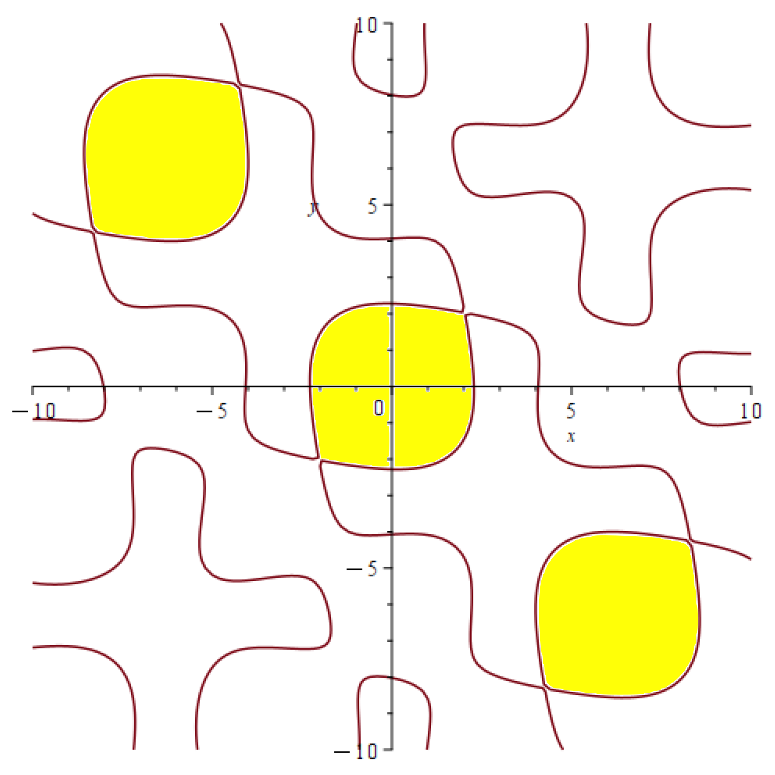}
\end{center}
\begin{center}
\includegraphics[width=0.8\linewidth]{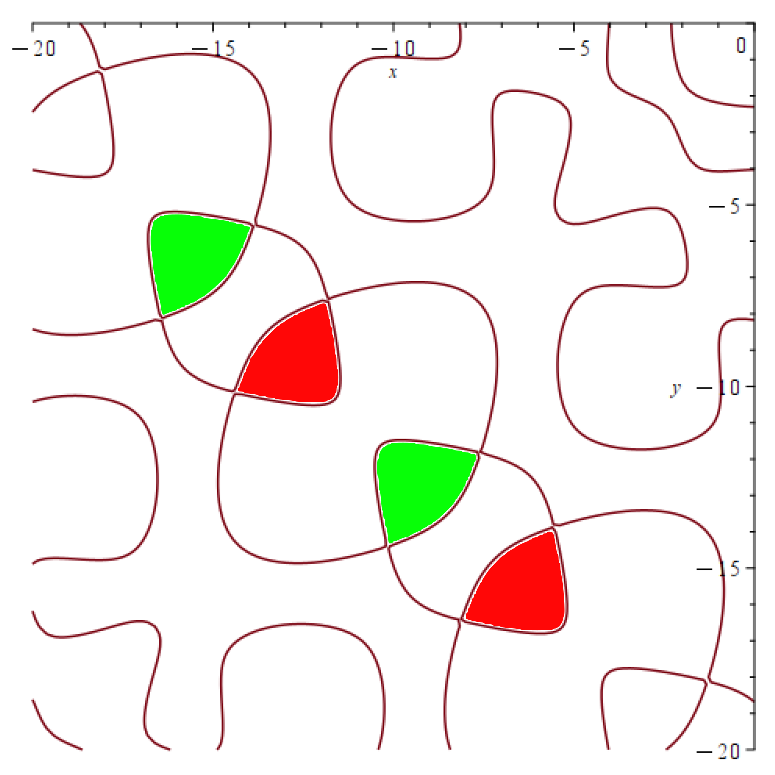}
\end{center}
\caption{Level lines (\ref{QLines1}) and (\ref{QLines2}) in the 
plane passing through the origin and in the shifted plane for 
$\, {\bf n} = Q \, $ and
$\, \epsilon_{F} \, = \, \widetilde{\epsilon}_{0} (Q) \, \simeq \, 
- 0.3502 \, $.
}
\label{PointQ}
\end{figure}

 As for the point $Q$, the heights of all three cylinders
$\, C_{1}, \, C_{2}, \, C_{3} \, $ also vanish at the level 
$\, \epsilon ({\bf p}) = \widetilde{\epsilon}_{0} (Q) \, $. 
In this case, however, the cylinders of zero height arise in 
different (parallel) planes of a given direction. 
Fig. \ref{PointQ} shows the plane passing through the origin, 
containing cylinders $\, C_{1} \, $ (of zero height), 
and a plane parallel to it, containing cylinders 
$\, C_{2} \, $ and $\, C_{3} \, $ (also of zero height). 
The corresponding level lines $\, \epsilon ({\bf p}) \, $ 
in these planes are given by the equations
\begin{multline}
\label{QLines1}
\cos x \, \cos y \,\,\, + \,\,\,(\cos x \, + \, \cos y) \,
\cos 0.2233 (x + y) \,\,\,  =   \\
=  \,\,\, - 0.3502
\end{multline}
and
\begin{multline}
\label{QLines2}
\cos x \, \cos y \,\,\, +  \\
+ \, (\cos x +  \cos y) \,
\cos \left( 0.2233 (x + y) + 0.198 \right) 
\,\, =  \,\, - 0.3502 \,\, , 
\end{multline}
which define the position of point $Q$ and the value
$\, \widetilde{\epsilon}_{0} (Q) = - 0.3502 \, $.
We thus obtain
$$\widetilde{\epsilon}_{0} (Q) \, - \,
\widetilde{\epsilon}_{0} (P) \,\, \simeq \,\,
0.009 \,\,\, , $$
which is approximately $\, 0.25 \% \, $ of the band width.

\vspace{1mm}

 As can be seen, the obtained estimate yields a non-zero, 
but rather small, value for the width of the interval 
$\, \left[ \epsilon^{\cal B}_{1} , \, \epsilon^{\cal B}_{2} \right] \, $, 
which indicates that it is indeed small for many real dispersion relations. 
In contrast to the two cases considered above, the emergence of chaotic 
trajectories in the tight-binding approximation for the face-centered lattice 
should be expected at a distance $\, \simeq \, 0.16 \, $ of the band width 
from the minimal energy in the conduction band. As in the previous two cases, 
here, apparently, it is also easier to achieve observation of the corresponding 
conductivity regimes using an external influence on the sample that regulates 
the position of the Fermi level.

\vspace{1cm}

\section{Conclusion}

 This paper examines the probability of detecting chaotic electron 
trajectories on the Fermi surface and the associated chaotic conductivity 
behavior in strong magnetic fields for dispersion relations obtained in 
the tight-binding approximation for cubic crystals. In crystals with the 
simple cubic or body-centered lattice, the energy range corresponding to the 
emergence of chaotic trajectories vanishes in the leading approximation. 
In crystals with the face-centered cubic lattice (the most common), this 
range is nonzero in the leading approximation but amounts to a fraction 
of a percent of the conduction band width. In all cases, observing chaotic 
regimes apparently requires controlling the position of the Fermi level 
by an external action. The position of chaotic trajectories differs 
significantly for the described cases: they appear in the center of the 
conduction band in the first two cases and are significantly shifted toward 
its lower edge in the latter case.

\end{document}